\documentclass{aa}  

\usepackage{graphicx}
\usepackage[varg]{txfonts}
\usepackage[T1]{fontenc}
\usepackage{nicefrac}
\usepackage{savesym}
\savesymbol{tablenum}
\usepackage[utf8]{inputenc}
\usepackage{amsmath}
\newcommand{\RN}[1]{%
  \textup{\uppercase\expandafter{\romannumeral#1}}%
}
\usepackage{mathtools}
\usepackage{booktabs}
\usepackage{multirow}
\usepackage[normalem]{ulem}
\usepackage[dvipsnames]{xcolor}
\usepackage{subcaption}
\usepackage{placeins}
\usepackage{enumitem}
\usepackage{tabularx}
\usepackage{siunitx}
\restoresymbol{SIX}{tablenum}
\DeclareSIUnit\parsec{pc}
\DeclareSIUnit\years{yr}
\DeclareSIUnit\Msol{\textit{M}_{\odot}}
\DeclareSIUnit\Lsol{\textit{L}_{\odot}}
\DeclareSIUnit\AU{au}
\DeclareSIUnit\om{\Omega}
\DeclareSIUnit\orb{T_{\mathrm{orb}}}
\DeclareSIUnit\scaleheight{H}
\newcommand{\Sigmag}{\Sigma_{\mathrm{g}}}

\newcommand{\rhog}{\rho_{\mathrm{g}}}
\newcommand{\rhod}{\rho_{\mathrm{d}}}
\newcommand{\vg}{\vec{v}_{\mathrm{g}}}

\newcommand{\amax}{a_{\mathrm{max}}}
\newcommand{\amin}{a_{\mathrm{min}}}
\newcommand{\aint}{a_{\mathrm{int}}}
\newcommand{\adr}{a_{\mathrm{drift}}}

\newcommand{\tfr}{t_\mathrm{fric}}
\newcommand{\St}{\mathrm{St}}
\newcommand{\Rey}{\mathrm{Re}}
\newcommand{\OmK}{\Omega_\text{K}}
\newcommand{\der}[2]{\frac{\partial {#1}}{\partial {#2}}}

\newcommand{\dpy}{\texttt{DustPy}}
\newcommand{\pluto}{\texttt{PLUTO}}
\newcommand{\tpop}{\texttt{TriPoD}}
\newcommand{\tpoppy}{\texttt{two-pop-py}}

\newcommand{\rev}[1]{{#1}}
\newcommand{\revI}[1]{{#1}}
\newcommand{\revII}[1]{{#1}}
\newcommand{\new}[1]{{#1}}

\makeatletter
\renewcommand*\aa@pageof{, page \thepage{} of \pageref*{LastPage}}
\makeatother

\usepackage{hyperref}
\usepackage[nameinlink, capitalise]{cleveref}
\creflabelformat{equation}{#2\textup{#1}#3}

\begin{document} 
   % \title{\tpop{}: A Dust Coagulation Sub-Grid Model for Hydrodynamic Simulations of Protoplanetary Disks}
   % \subtitle{Dust Coagulation in Vertically-Integrated Simulations with the \pluto{} Code}

    \title{\tpop{}: \texttt{Tri}-\texttt{P}opulation size distributions for \texttt{D}ust evolution.}
   \subtitle{Coagulation in vertically integrated hydrodynamic simulations of \\ protoplanetary disks.}

   \author
   {
        Thomas Pfeil \inst{123_{\href{https://orcid.org/0000-0002-4171-7302}{\includegraphics[height=1.2em]{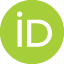}}}}
        \and
        Til Birnstiel \inst{14_{\href{https://orcid.org/0000-0002-1899-8783}{\includegraphics[height=1.2em]{orcid.png}}}}
        \and
        Hubert Klahr \inst{2_{\href{https://orcid.org/0000-0002-8227-5467}{\includegraphics[height=1.2em]{orcid.png}}}}
    }

   \institute
   {
        University Observatory, Faculty of Physics, Ludwig-Maximilians-Universität München, Scheinerstr. 1, D-81679 Munich, Germany
        \and 
        Max-Planck-Institut f\"ur Astronomie, K\"onigstuhl 17, D-69117 Heidelberg, Germany
        \and  Center for Computational Astrophysics, Flatiron Institute, 162 Fifth Avenue, New York, NY 10010, USA \\
        \email{tpfeil@flatironinstitute.org}
        \and Exzellenzcluster ORIGINS, Boltzmannstr. 2, D-85748 Garching, Germany
    }
    
   % \date{Received September 15, 1996; accepted March 16, 1997}

% \abstract{}{}{}{}{} 
% 5 {} token are mandatory
 
  \abstract
  % context heading (optional) 
   {Dust coagulation and fragmentation impact the structure and evolution of protoplanetary disks and set the initial conditions for planet formation. Dust grains dominate the opacities, they determine the cooling times of the gas via thermal accommodation in collisions, they influence the ionization state of the gas, and the available grain surface area is an important parameter for the chemistry in protoplanetary disks. Therefore, dust evolution is an effect that should not be ignored in numerical studies of protoplanetary disks. Available dust coagulation models are, however, too computationally expensive to be implemented in large-scale hydrodynamic simulations. This limits detailed numerical studies of protoplanetary disks, including these effects, mostly to one-dimensional models.}
  % aims heading (mandatory)
   {We aim to develop a simple---yet accurate---dust coagulation model that can be easily implemented in hydrodynamic simulations of protoplanetary disks. Our model shall not significantly increase the computational cost of simulations and provide information about the local grain size distribution.}
  % methods heading (mandatory)
   {The local dust size distributions are assumed to be truncated power laws. Such distributions can be fully characterized by only two dust fluids (large and small grains) and a maximum particle size, truncating the power law. We compare our model to state-of-the-art dust coagulation simulations and calibrate it to achieve a good fit with these sophisticated numerical methods.}
  % results heading (mandatory)
   {Running various parameter studies, we achieved a good fit between our simplified three-parameter model and \dpy{}, a state-of-the-art dust coagulation software.}
  % conclusions heading (optional), leave it empty if necessary 
   {We present \tpop{}, a sub-grid dust coagulation model for the \pluto{} code. With \tpop{}, we can perform two-dimensional, vertically integrated dust coagulation simulations on top of a hydrodynamic simulation. Studying the dust distributions in two-dimensional vortices and planet-disk systems is thus made possible.}

   \keywords{protoplanetary disks --- dust evolution --- hydrodynamics --- methods: numerical}

   \maketitle
%
%-------------------------------------------------------------------

\section{Introduction} \label{sec:intro}
Models of dust coagulation in protoplanetary disks are required to understand the formation of \si{\centi \meter}-sized pebbles \citep{Brauer2008, Birnstiel2009} and \si{\kilo \meter}-sized planetesimals \rev{\citep{Wetherill1989, Schlichting2011, Kobayashi2016,Lau2022, Drazkowska2022}}; they are indispensable for the interpretation of observational data \citep{Birnstiel2018, Dullemond2018} and necessary to simulate the assembly of whole planetary systems \citep{Lichtenberg2021, Emsenhuber2021}. 
The size of dust grains also determines their aerodynamic properties and thus sets the timescales at which grains drift towards the central star or collect in local pressure maxima \citep{Whipple1972, Weidenschilling1977}.
Furthermore, dust is the dominating source of opacity in circumstellar disks, which means the size distribution of the dust grains has a strong influence on the disks' thermal structure \citep{Muley2023} and hydrodynamics \citep{Lesur2022}, as well as on the interpretation of observations \citep{Birnstiel2018,Leiendecker2022,Bergez-Casalou2022, Antonellini2023}. In addition, the presence of small grains sets limits to the disks' ionization and is thus also important for studies of magnetohydrodynamic mechanisms like the MRI \citep{Balbus1991} and magnetized disk winds \citep{Blandford1982} with non-ideal MHD effects \citep{Guillet2020, Pascucci2022, Tsukamoto2022}.
In 2018, observation with the Atacama Large Millimeter/submillimeter Array (ALMA) revealed that a broad variety of substructures exist in the spatial distribution of dust in protoplanetary disks \citep{Andrews2018}. Numerous gaps, spirals, and vortices have since been observed in the dust continuum emission \citep[e.g.,][]{Perez2018, Baruteau2019, Tsukagoshi2022} and also in molecular line observations \citep[see][]{Oberg2021}.
The existence of these structures raises questions regarding their origins and how they impact the formation and composition of planetesimals and planets within the disks---which makes hydrodynamic simulations of protoplanetary disks, including dust coagulation models, necessary \rev{\citep{Birnstiel2018, Drazkowska2019}}.

\begin{figure*}[t]
    \centering
    \includegraphics[width=0.99\textwidth]{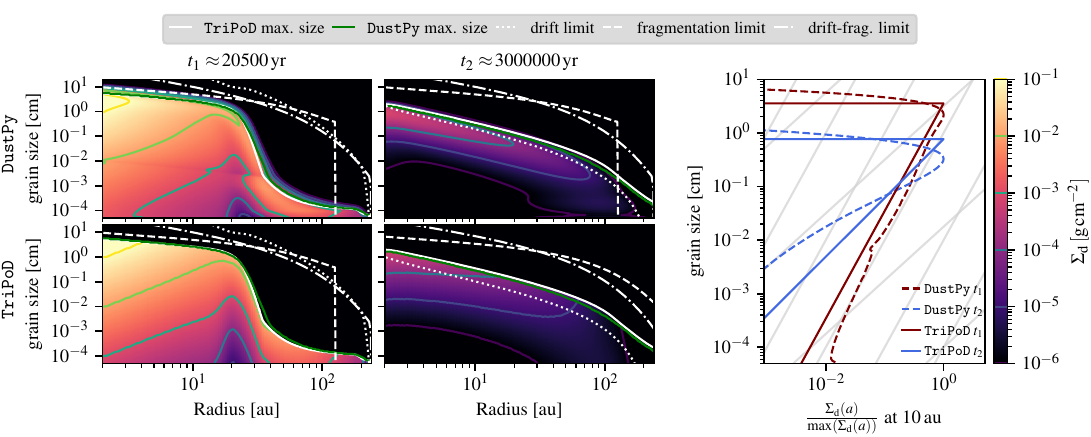}
    \caption[Comparison between the fiducial \dpy{} setup and the fiducial \tpop{} setup.]{Comparison between the full coagulation model \dpy{} (upper row, 171 dust fluids) and our new three-parameter power-law prescription \tpop{}, which we implemented in the \pluto{} code (two dust fluids). We show two snapshots of the one-dimensional disk models in the first two panels in each row. The third panel shows the local dust size distribution of the respective model at \SI{10}{\AU} for both snapshots. The light grid in the background represents size distribution power laws with $n(a)\propto a^{-3.5}$ and $n(a)\propto a^{-3.0}$.}
    \label{fig:fiducial}
\end{figure*}

Modeling the evolution of the solid content of protoplanetary disks has proven to be an expensive task in computational astrophysics. Many dust evolution models \citep[e.g.][]{Nakagawa1981, Weidenschilling1980, Brauer2008, Birnstiel2010, Charnoz2012, Drazkowska2019}, numerically solve a discretized form of the Smoluchowski coagulation equation \citep{Smoluchowski1916}.
The coagulation equation's integro-differential nature makes solving it, however, numerically costly. This procedure utilizes a grid of grain sizes, meaning that dozens or hundreds of dust fluids have to be modeled in a single simulation, each representing grains of a different size that interact with every other grain size via collisions. Studies of dust coagulation are often carried out in one-dimensional disk models \rev{in either the radial direction} \citep[e.g.,][]{Lenz2020, Drazkowska2021, Pinilla2021, Garate2021, Burn2022}, \rev{the vertical direction \citep[e.g,][]{Zsom2011, Krijt2016}}, or limited two-dimensional studies \citep{Drazkowska2019}. 

Therefore, efforts are pursued to solve the coagulation equations more efficiently, such as the use of new \rev{numerical} methods \revII{\citep[e.g.,][]{Estrada2008,Lombart2021,Lombart2022,Lombart2024,Marchand2021,Marchand2022,Marchand2023}}, or by applying far-reaching simplifications to the physics of dust coagulation \citep{Birnstiel2012} that make it possible to implement dust coagulation as a subgrid model in hydrodynamic simulations \citep{Tamfal2018, Vorobyov2018}. Machine-learning-aided techniques also promise a fast, yet simplified approach to model dust coagulation on top of a hydrodynamic simulation \citep{Pfeil2022}.

Here, we present a semi-analytic model of dust coagulation, which is based on a two-population approach, \rev{originally developed by \citep{Birnstiel2012} and employed in various forms by others \citep{Tamfal2018, Vorobyov2019, Vorobyov2019a, Vorobyov2020}. This model, however, has the critical disadvantage of not evolving the full dust size distribution but only the maximum particle size.} With our new model, we are now able to conduct two-dimensional, vertically integrated hydrodynamics simulations of protoplanetary disks with an evolving dust size distribution at low computational cost.
\Cref{fig:fiducial} shows an example comparison between a full coagulation simulation, conducted with \dpy{} \citep{Stammler2022}---a full-fledged dust coagulation software---and a one-dimensional hydrodynamic simulation with the \pluto{} code equipped with our new method.
While one-dimensional hydrodynamics simulations do not allow for a direct comparison of performance due to the different methods to handle the transport (\dpy{} utilizes an implicit integration scheme \rev{and does not solve the equations of hydrodynamics, but an advection-diffusion equation for a Keplerian disk}), they allow for detailed tests of the accuracy of our new model.

This article is structured as follows:
In \cref{sec:theory}, we briefly review the physics of dust dynamics and coagulation. We also give a short description of \tpoppy{} \citep{Birnstiel2012}, the progenitor of our new model. \Cref{sec:model} introduces our new three-parameter dust coagulation model \tpop{} and how it is integrated in the \pluto{} hydrodynamics code \citep{Mignone2007}. As \tpop{} is a highly simplified prescription of dust coagulation, we have to calibrate it to achieve a good fit with full-coagulation models. The respective calibration runs are presented in \cref{sec:calibration}. In \cref{sec:tests}, we present test simulations that demonstrate the accuracy of \tpop{} in comparison with \dpy{} simulations. We also give an example of a two-dimensional simulation of a planet-disk system in which we compare the outcome of our new model to the old \tpoppy{} model.
We discuss the limitations of our approach in \cref{sec:discussion} and summarize in \cref{sec:summary}.

\section{Theory}
\label{sec:theory}
\subsection{Dust-gas relative motion}
Dust particles moving in the gaseous protoplanetary disk can experience drag forces due to differences between the equilibrium velocities of dust and gas particles. The gas in a PPD is radially stratified, which means the disk has a radial pressure gradient. Consequently, gas moves on a slightly sub-Keplerian orbit, where hydrostatic equilibrium is given by
\begin{equation}
\Omega^2 =\frac{1}{R \rhog}\frac{\partial P}{\partial R} + \OmK^2\, , 
\end{equation} 
where $P$ and $\rhog$ are the gas pressure and density respectively, $R$ is the cylindrical, stellocentric radius, and $\OmK^2 = \nicefrac{GM_*}{R^3}$ is the Keplerian angular frequency \revII{(see \cref{tab:Sym} for a list of symbols)}. 
Conversely, \rev{radial pressure forces are negligible for the solid particles}, which means their equilibrium orbits would be Keplerian (ignoring the gas drag and additional effects like radiation pressure, etc.).
Gas, at velocity $\vec{v}_\mathrm{g}$ and dust particles, at velocity $\Vec{v}_\mathrm{d}$ are aerodynamically coupled via a friction force density
\begin{equation}
    \vec{f}_\text{fric} = \rhod\frac{\vec{v}_\mathrm{d}-\vec{v}_\mathrm{g}}{\tfr}\, , 
\end{equation}
where the strength of the coupling can be characterized by the stopping timescale $\tfr$. In the Epstein regime \citep{Epstein1924}, the friction time can be written
\begin{equation}
    \tfr = \sqrt{\frac{\pi}{8}}\frac{\rho_\text{m} a}{\rhog c_\mathrm{s}}\, ,
\end{equation}
where $\rho_\mathrm{m}$ is the particles' material density, $a$ denotes the particle radius, and $c_\mathrm{s}$ is the sound speed.
A useful dimensionless measure of the strength of the coupling between gas and dust particles is the Stokes number $\St \coloneqq \tfr\OmK$. For $\St\ll 1$, gas and dust particles are well-coupled and the particles quickly adjust to changes in the gas velocity. For $\St\gg1$ however, particles are decoupled from the gas and the friction force is no longer significant for the trajectories of the dust grains.

The stopping time is the timescale on which the particles and the gas \rev{approach} a steady state. The respective terminal velocities of the grains in force equilibrium were derived by \cite{Nakagawa1986}. \rev{Ignoring additional velocity components of the gas, for example, due to viscous evolution, }the respective relative \rev{radial} velocity between the grains and the gas then follows as
\begin{align}
    v_{\text{d-g},R} &= \frac{\St (1+\varepsilon)}{\St^2 + (1+\varepsilon)^2}\frac{1}{\OmK\rhog}\frac{\partial P}{\partial R} 
    \\&\approx \frac{\St}{\St^2+1}\frac{1}{\OmK\rhog}\frac{\partial P}{\partial R} \quad \text{for} \quad \varepsilon \ll 1\, ,
    \label{eq:driftvel}
\end{align}
where $\varepsilon$ is the dust-to-gas density ratio. Thus particles drift towards pressure maxima and reach their maximum terminal velocity at a Stokes number of one.

\subsection{Dust-dust relative motion}
Relative velocities between the gas and dust depend on the aerodynamic properties of the dust. Differently sized grains therefore experience relative velocities due to gas drag. 
For small grains, Brownian motion is of importance. Additionally, turbulence causes random variations in the gas velocities that act on the dust particles according to their aerodynamic coupling to differently sized eddies in the gas.

\paragraph{Brownian motion}
Random molecular motion of particles leads to relative velocities that depend on the respective particles' masses \rev{\citep{Brauer2008}}
\begin{equation}\label{eq:Brown}
    \Delta v_\text{01\,Brown} = \sqrt{\frac{8k_\text{B}T(m_0+m_1)}{\pi m_0 m_1}}\, ,
\end{equation}
where $m_0$ and $m_1$ denote the particles' masses, $T$ is the gas temperature, and $k_\mathrm{B}$ is the Boltzmann constant.
This effect is only relevant for the smallest particles on \rev{micrometer} scales.

\paragraph{Relative drift velocities} 
For two particles with Stokes numbers $\St_0$ and $\St_1$, the relative drift velocities in the case of low dust-to-gas ratio, are given by \cref{eq:driftvel}
\begin{equation}
    \Delta v_\text{01\,drift} = \left| \left( \frac{\St_0}{\St^2_0+1}-\frac{\St_1}{\St^2_1+1}\right)\frac{1}{ \OmK\rhog} \frac{\partial P}{\partial R}\right|\, .
\end{equation}

\paragraph{Relative settling velocities}
\citet{Dubrulle1995} derived the vertical dust distribution in a disk with turbulent diffusion as
\begin{equation}
    \rhod(z)=\rho_\mathrm{d,mid}\exp\left(-\frac{z^2}{2  H_\mathrm{d}}\right)\, ,
\end{equation}
where $H_\text{d}$ is the scale height of the dust, given by
\begin{equation}
    H_\mathrm{d}=\frac{H}{\sqrt{1+\frac{\St}{\delta}}}\, ,
\end{equation}
\rev{\citep[see also][]{Fromang2009, Binkert2023}}.
Here, $H$ refers to the gas scale height, and $\delta$ denotes the turbulent diffusivity \rev{parameter}, which for now is assumed to be equal to the turbulent gas viscosity parameter $\alpha$.
Thus, particles of different sizes are, on average, also found at different heights, and thus have different terminal velocities. 
The average relative velocities between the two particle populations can then be approximated as 
\begin{equation}\label{eq:rel_set}
    \Delta v_\text{01\,set} = |H_{\text{d}1} \St_1 - H_{\text{d}0}\St_0|\, \OmK \, .
\end{equation}

\paragraph{Relative velocities due to turbulence}
\citet{Ormel2007} derived closed-form expressions for the relative particle velocities in different turbulence regimes, which depend on the Stokes numbers, the turbulent gas velocity, and the local Reynolds number
\begin{align}
\Rey &= \frac{\nu_{\text{turb}}}{\nu_{\text{mol}}} \approx \frac{\alpha c_\mathrm{s} H}{c_\mathrm{s} \lambda_{\text{mfp}}} = \frac{\alpha H \rhog \sigma_{\mathrm{H}_2}}{\mu m_\mathrm{p}}\, ,
\end{align}
where $\lambda_\mathrm{mfp}$ is the mean free path of the gas molecules, $\mu m_\mathrm{p}$ is the mean molecular mass of the gas, and $\sigma_{\mathrm{H}_2}$ is the collisional cross-section of two gas molecules (here $\mathrm{H_2}$).
The respective derivations assume a Kolmogorov turbulent energy cascade \citep{Kolmogorov1941}.
We use an implementation of these velocities identical to the one utilized in the full dust coagulation code \dpy{}.

\subsection{Dust coagulation}
\label{sec:dustcoag}
Dust particles in protoplanetary disks undergo collision since they experience differential velocities due to their interaction with the gas.
\rev{Surface} forces can lead to sticking in such collisions and thus facilitate the growth of dust particles.
If collision velocities are too high, they can lead to fragmentation. The Smoluchowski equation \citep{Smoluchowski1916} describes the evolution of continuous mass (or size) distributions of grains $n(m)$, as a consequence of these processes.
In this work, however, we are not numerically solving the coagulation equation as in full-fledged coagulation models like \dpy{} \citep{Stammler2022}. Instead, we use the results gained with such elaborate numerical methods to construct a simplified, semi-analytic prescription for dust coagulation. For this, it is instructive to have a look at some main results obtained with full coagulation models and simple analytic derivations.
One of these simplifying assumptions is a monodisperse size distribution. \cite{Kornet2001} have shown that for such a case, the particle growth rate can be written as
\begin{equation}\label{eq:monogr}
    \dot{a}=\frac{\rhod}{\rho_\text{m}}\Delta v,
\end{equation}
where $\Delta v$ denotes the relative velocity between the grains. If the relative velocities are assumed to be caused by gas turbulence in the fully intermediate regime, one finds that the growth of the particles is occurring on a timescale 
\begin{equation}
    t_\mathrm{grow}=\frac{1}{\varepsilon \Omega_\mathrm{K}}\, .
\end{equation}
However, dust grains can only grow in size as long as their relative velocities due to different aerodynamic properties are not above a critical fragmentation velocity $v_\mathrm{frag}$, for which collisions would lead to the destruction of the particles. \rev{Another possible outcome of grain collisions is bouncing, as shown by laboratory experiments \citep{Guttler2010} and studied in numerical models \citep{Zsom2010, Dominik2024}.}
It is possible to derive analytic estimates for the maximum reachable particle size, given a certain fragmentation velocity.
\citet{Birnstiel2012} derived the respective maximum particle size in the turbulent fragmentation limit as
\begin{equation}\label{eq:frag_limit}
    a_{\mathrm{turb\text{-}frag}} = \sqrt{\frac{8}{\pi}} \frac{\rhog}{3\rho_{\mathrm{m}}} \frac{v_\mathrm{frag}^2}{\alpha c_\mathrm{s} \OmK}\, .
\end{equation}
Furthermore, dust grains undergo radial drift, which results in relative velocities between grains of different sizes (see \cref{eq:driftvel}). The resulting collisions can also lead to a drift-fragmentation limit, which is given by
\begin{equation}\label{eq:dr_fr_limit}
      a_{\mathrm{drift\text{-}frag}} =\sqrt{\frac{8}{\pi}} \frac{\rhog}{\rho_{\mathrm{m}}} \frac{v_{\mathrm{frag}}}{c_\mathrm{s}} \frac{P}{1-\mathcal{N}} \left| \der{P}{r} \right| ^{-1}\, , 
\end{equation}
where the constant $\mathcal{N}$ is approximately 0.5 \citep{Birnstiel2012}.
Dust particles can thus not reach sizes larger than
\begin{equation}
    a_\mathrm{frag} = \mathrm{min}(a_{\mathrm{drift\text{-}frag}}, a_{\mathrm{turb\text{-}frag}})\, .
\end{equation}

\begin{figure}[t]
    \centering
    \includegraphics[width=\columnwidth]{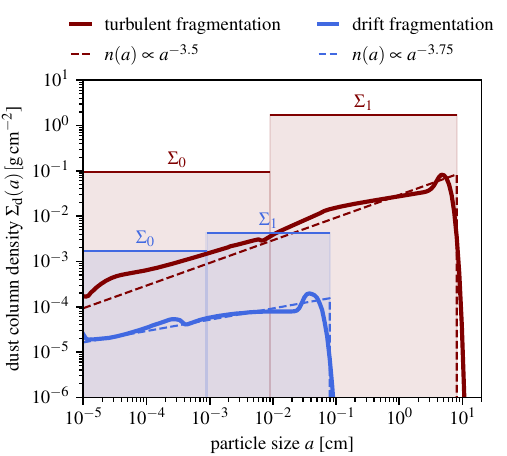} 
    \caption[Typical dust size distributions and their power law equivalents.]{Typical size distributions in regions of a protoplanetary disk model that are either turbulence-dominated (dark red lines) or drift-dominated (blue lines). Both regimes result in distinct power-law exponents of the distribution that we overplot as dashed lines. Assuming a power law as the overall shape of the distribution makes it possible to approximately describe it with only three parameters: The cutoff particle size $\amax$ and the two densities $\Sigma_0$ (contained in the size interval $[\amin,\,\sqrt{\amax \amin}$) and $\Sigma_1$ (contained in the size interval $[\sqrt{\amax \amin},\,\amax]$). This approximation is the basis of our new dust coagulation model.}
    \label{fig:fragdistr}
\end{figure}

In these cases, the typically reached size distributions are approximately power laws with characteristic exponents. \cite{Birnstiel2011} used coagulation simulations and analytical calculations to find these fragmentation-limited size distributions. 
They derived analytic expressions for the resulting power-law exponent of a mass distribution $n(m)\propto m^{-q_\mathrm{m}}$ in three different regimes, translating to the size distribution $n(a)\propto a^q = a^{-3q_\mathrm{m}+2}$. If coagulation and fragmentation happen simultaneously, the power-law exponent can be written as
\begin{equation}
    q=2-\frac{3}{2}(\nu+\xi+1)\, ,
\end{equation}
where $\nu$ is a kernel index that depends on how the relative velocities change with particle size. The parameter $\xi$ determines the typical distribution of fragments in a destructive collision. It is usually set to the canonical value of $\nicefrac{11}{6}$. 
\cite{Birnstiel2011} determined that in the first regime of turbulence, as derived by \cite{Ormel2007}, where the relative velocities scale linearly with particle size, $\nu=1$ and thus $q=-3.75$. The same is true for relative drift velocities, which also linearly depend on the difference in Stokes numbers between the colliding particles (as long as $\St\ll 1$).
This means that typical size distributions in coagulation-fragmentation equilibrium, with collisions driven by differential drift or the first regime of turbulence, can be approximated as power laws $n(a)\propto a^{-3.75}$, as illustrated in \cref{fig:fragdistr}.
This case is thus relevant whenever the particle fragmentation velocity is low and only small particles exist, or if the drift velocities dominate over the turbulent velocities, as in the outer regions of protoplanetary disks.

The other prominent case is relevant whenever turbulence is causing relative velocities between particles in the so-called fully intermediate regime.
Then, $\nu=\nicefrac{5}{6}$, and one finds that $q=-3.5$, which is equal to the \cite*{Mathis1977} (MRN) distribution. This particular case is relevant in the inner parts of protoplanetary disks, where the particles grow to the largest sizes and where radial drift is less relevant. 

Finally, we have the stages of dust growth in which the fragmentation barrier is not yet reached and the largest particles are undergoing a sweep-up growth. In this case, typical size distributions are steeper and we assume $q=-3.0$ in this case \rev{\citep{Simon2022, Birnstiel2023}}.

Radial drift itself also sets a limit to the maximum particle size, which is reached when the radial drift time scale becomes equal to the local growth time scale
\begin{align}\label{eq:dr_limit}
    \frac{\adr}{\dot{a}} &\stackrel{!}{=} \frac{R}{|\dot{R}|} \nonumber \\
    \adr &= \sqrt{\frac{8}{\pi}} \frac{\rhod}{\rho_{\mathrm{m}}} \frac{v_{\mathrm{K}}}{c_\mathrm{s} \gamma}\, ,
\end{align}
where $v_\mathrm{K}$ is the Keplerian velocity and $\gamma=\left|\frac{\mathrm{d}\log P}{\mathrm{d}\log R}\right|$ denotes the radial double-logarithmic pressure gradient. 
The drift limit is relevant in the outer regions of protoplanetary disks, where drift can become rapid for large particles. In the drift limit, grains do not undergo fragmentation. Therefore, typical size distributions are sweep-up dominated and accordingly steep. We assume $q=-3.0$, which is typically a good fit to full coagulation simulations.

\subsection[The two-pop-py model by Birnstiel et al.\ (2012)]{The \tpoppy{} model by \citep{Birnstiel2012}}
With \tpoppy{}, \cite{Birnstiel2012} introduced a strongly simplified and very fast method to model the effects of dust coagulation in protoplanetary disks. 
In this method, dust is realized as a single fluid that drifts relative to the gas. The flux calculation, however, considers two dust species. The small population represents the monomers with fixed size $a_\mathrm{mon}$ and is assumed to move along with the gas. Larger, grown grains make up the population of size $a_\mathrm{gr}(t)$, which is evolving in time, following \rev{the monodisperse dust growth rate, which is limited by the above-discussed growth barriers}
\begin{equation}
\label{eq:tpoppysize}
    a_\mathrm{gr}(t) = \min\left[\min(a_\mathrm{drift},\, a_\mathrm{turb\text{-}frag},\, a_\mathrm{drift\text{-}frag}),\, a_\mathrm{mon}\exp\left(\frac{t-t_0}{t_\mathrm{grow}}\right)\right]\, ,
\end{equation}
to simulate the initial phase of coagulation and the growth limits.
Both species are associated with a drift velocity according to their Stokes numbers $\St_\mathrm{mon/gr}=\nicefrac{a_\mathrm{mon/gr}\rho_\mathrm{m}\pi}{2\Sigma_\mathrm{g}}$, as
\begin{equation}
    v_\mathrm{mon/gr}=\frac{v_\mathrm{gas}}{1+\St_\mathrm{mon/gr}^2} + \frac{\St_\mathrm{mon/gr}}{1+\St_\mathrm{mon/gr}^2}\frac{1}{\rhog \Omega_K}\frac{\partial P}{\partial r}\, ,
\end{equation}
where the first term takes into account gas velocities arising from viscous disk evolution and the second term is the radial drift velocity derived by \cite{Nakagawa1986} in the limit of small dust-to-gas ratios.
The total dust flux velocity follows as a mass average of both populations, via
\begin{equation}
    \Bar{v}=(1-f_\mathrm{m})v_\mathrm{mon} + f_\mathrm{m}v_\mathrm{gr}\, .
\end{equation}
The ratio $f_\mathrm{m}$ is dependent on the limiting factor of grain growth: fragmentation-limited, drift-fragmentation-limited, drift-limited, or neither when the dust is still in the growth phase
\begin{equation}
    f_\mathrm{m} = \begin{cases}
        0.97 \,\, \text{if }a_\mathrm{drift} < a_\mathrm{frag},\, a_\mathrm{drift\text{-}frag} \\
        0.75 \,\, \text{otherwise.}
    \end{cases}
\end{equation}
Given the resulting velocity, a flux can be calculated that is used to evolve the dust surface density in time.
This method is fast and can be easily implemented in a hydrodynamics code. It has however some serious drawbacks:
\begin{itemize}
    \item Dust growth is always assumed to be limited by an equilibrium of either fragmentation and coagulation or drift and coagulation. This is mostly true if no substructure is present. In some situations like planetary gaps, however, fluxes and grain sizes are no longer determined by coagulation in the gap but by the supply of small grains that diffuse into the gap and the efficient removal of larger grains. In such cases, grain sizes can be underestimated by \tpoppy{}, which would assume the drift limit.
    \item Although the maximum grain size is known, no information on the actual size distribution is provided \rev{beyond the knowledge of whether the distribution is drift-limited or fragmentation-limited}. 
    \item The differences between the size distribution in the fragmentation limit and the drift-fragmentation limit are not taken into account. The mass fraction $f_\mathrm{m}$ only considers whether the maximum particle size is drift-limited.
    \rev{
    \item As growth is always assumed to be driven by collisions in the fully intermediate regime of turbulence, the actual growth timescale can be underestimated \citep{Powell2019}.
    \item The fragmentation limit is reached instantaneously and not gradually. This is especially problematic if the timescale for dust advection is short.
    \item The model is only calibrated to reproduce the dust size evolution in protoplanetary disks around \SI{1}{\Msol} stars. 
    }
    \item The model overestimates the concentration of dust in pressure bumps due to the treatment of dust being transported as one fluid. Although the flux is calculated in a mass-averaged way that has been determined experimentally, intermediately sized grains are neglected, which would lead to a wider dust distribution in pressure bumps.
\end{itemize}
With our new model, \tpop{}, we aim to mitigate these problems.

\section{The \tpop{} model}
\label{sec:model}
Our new three-parameter dust evolution model \tpop{} makes use of the knowledge gained from full-fledged coagulation models that are discussed in the previous section.
In particular, the power-law prescriptions of the dust size distribution are the basis on which we build our method. \rev{\tpop{} describes the power-law size distribution with only three parameters: the dust column densities of a small population $\Sigma_0$ and the large population $\Sigma_1$, as well as a maximum particle size $\amax$ at which the distribution is truncated \revII{(see \cref{tab:Sym} for a list of symbols used in the context of \tpop{})}. The populations are separated at the size $\aint=\sqrt{\amin\amax}$, which then defines the power-law exponent of a full size distribution via
\begin{equation}\label{eq:2popdistr}
    q = \frac{\log(\Sigma_1/\Sigma_0)}{\log(\amax/\aint)} - 4\, .
\end{equation}
Note, that this formula is in general independent of whether we describe a size distribution in the sense of column densities or volume densities. Although we develop the \tpop{} method in this paper for use in vertically integrated size distributions, we could also do so for volume densities. In the following, we will thus oftentimes use the dust-to-gas ratio $\varepsilon$, which can be interpreted as either the vertically integrated, or the local version. 
}
Assuming the dust size distribution to follow a truncated power law $n(a)\propto a^{q}$, we can write the dust-to-gas ratio size distribution as $\varepsilon(a) \propto a^{q}m\propto a^q a^3$. Normalizing this to the total dust-to-gas ratio $\varepsilon_\mathrm{tot}$, we get
\begin{align}
   \varepsilon(a)= \begin{cases}
    %\frac{\varepsilon_{\text{tot}}\amax\amin}{\amax-\amin}\frac{1}{a^2} & \text{for } p=-5 \\
    \dfrac{\varepsilon_{\text{tot}}(q+4)}{\amax^{q+4} - \amin^{q+4}} a^{q+3} & \text{for } q\neq -4 \\[15pt]
    \dfrac{\varepsilon_{\text{tot}}}{\log(\amax)-\log(\amin)}\dfrac{1}{a} & \text{for } q=-4\, ,
  \end{cases}
\end{align}
where $\amin$ is the minimum particle size and $\amax$ is the maximum particle size of the truncated distribution.
From this, the dust-to-gas ratio within a given size interval $[a_\RN{1}, a_\RN{2}]$ with $\amin\leq a_\RN{1} < a_\RN{2} \leq \amax$, can be calculated as
\begin{align}
\begin{split}
        \varepsilon_{a_\RN{1}}^{a_\RN{2}} &\coloneqq \int_{a_\RN{1}}^{a_\RN{2}} \varepsilon(a)\,\mathrm{d}a \\
    &=  \begin{cases}
    %\frac{\varepsilon_{\text{tot}}\amax\amin}{\amax-\amin}\frac{1}{a^2} & \text{for } p=-5 \\
    \varepsilon_{\text{tot}}\dfrac{a_\RN{2}^{q+4}-a_\RN{1}^{q+4}}{\amax^{q+4}-\amin^{q+4}} & \text{for } q\neq -4 \\[15pt]
    \varepsilon_{\text{tot}}\dfrac{\log(a_\RN{2})-\log(a_\RN{1})}{\log(\amax)-\log(\amin)} & \text{for } q=-4\, .
  \end{cases}    \label{eq:distr}
\end{split}
\end{align}
Similarly, the mass-averaged particle size of this distribution is defined as
\begin{align}
\begin{split}
    \langle a\rangle_{a_\RN{1}}^{a_\RN{2}} &\coloneqq  \dfrac{\int_{a_\RN{1}}^{a_\RN{2}} \varepsilon(a)\,a\,\mathrm{d}a}{\int_{a_\RN{1}}^{a_\RN{2}} \varepsilon(a)\, \mathrm{d}a}  \\
&= \begin{cases}
    \dfrac{a_\RN{2}a_\RN{1}}{a_\RN{2}-a_\RN{1}}\log\left(\dfrac{a_\RN{2}}{a_\RN{1}}\right) & \text{for } q=-5 \\[15pt]
    \dfrac{q+4}{q+5} \dfrac{a_\RN{2}^{q+5} - a_\RN{1}^{q+5}}{a_\RN{2}^{q+4} - a_\RN{1}^{q+4}} & \text{for } q\neq -5,-4 \\[15pt]
    \dfrac{a_\RN{2}-a_\RN{1}}{\log(a_\RN{2})-\log(a_\RN{1})} & \text{for } q=-4\, .
  \end{cases} \label{eq:amean}
\end{split}
\end{align}
In the \tpop{} model, we define two particle populations that together contain the entire dust density of the distribution
\begin{align}
\begin{split}
     \varepsilon_{0} &= \int_{a_{\mathrm{min}}}^{a_{\mathrm{int}}} \varepsilon(a)\,\mathrm{d}a \\ \varepsilon_{1} &= \int_{a_{\mathrm{int}}}^{a_{\mathrm{max}}} \varepsilon(a)\,\mathrm{d}a\, .
\end{split}
\end{align}
\rev{The mass-averaged particle sizes of both populations are then given by $a_0\coloneqq\langle a\rangle_{\amin}^{\aint}$ and $a_1\coloneqq\langle a\rangle_{\aint}^{\amax}$ (see \cref{eq:amean}).}
It can be shown that the two populations $\varepsilon_0$ and $\varepsilon_1$ exactly represent the power-law distribution if $\aint$ is defined as the geometric mean of the maximum and minimum size $\aint=\sqrt{a_\text{max}\amin}$. 
The power-law exponent $q$ is then given by \cref{eq:2popdistr}.
Knowing only $\varepsilon_0$ and $\varepsilon_1$ and the maximum size $\amax$ thus allows us to reconstruct the entire size distribution. 

On a size grid with $N$ cells $a_i$, spanning from $\amin$ to $a_N\geq\amax$, we can write the mass in a single size bin, and likewise the entire size distribution as
% \begin{equation}
%     \Sigma_{\text{d},i} = \Sigma_\text{d} \frac{a_i^{q+4}\Theta[a_\text{max}-a_i]}{\sum_{i=0}^N a^{q+4}_i\Theta[a_\text{max}-a_i]},
% \end{equation}
\begin{equation}
\begin{split}
    \varepsilon_{i} =
    \begin{cases}
        \varepsilon_\mathrm{tot} \dfrac{a_{i+\nicefrac{1}{2}}^{q+4}-a_{i-\nicefrac{1}{2}}^{q+4}}{\amax^{q+4}-\amin^{q+4}} \theta(a) & \text{for } q\neq -4 \\[15pt]
        \varepsilon_\mathrm{tot} \dfrac{\log(a_{i+\nicefrac{1}{2}})-\log(a_{i-\nicefrac{1}{2}})}{\log(\amax)-\log(\amin)} \theta(a)& \text{for } q=-4\, ,
    \end{cases} 
\end{split}
\end{equation}
where $a_{i-\nicefrac{1}{2}}$ and $a_{i+\nicefrac{1}{2}}$ denote the cell interfaces on the size grid, $\varepsilon_\mathrm{tot}=\sum_{i=0}^N \varepsilon_{i}$ is the total dust-to-gas density ratio and $\theta(a)=\Theta(a_i-\amin)\Theta(a_i-\amax)$ represents two Heaviside step functions that cut off the distribution at the minimum and maximum particle sizes. 
This means we can directly compare our model results with full dust coagulation models like \dpy{}, which evolve a large grid of sizes instead of just two fluids in our case. We illustrate this in \cref{fig:fragdistr}, where we overplot the detailed size distributions, obtained with a full coagulation model, with their respective three-parameter size distribution representation. The respective dust column densities $\Sigma_0$ and $\Sigma_1$ of the two populations are shown as the horizontal bars spanning the respective size ranges.

In the following, we describe how we evolve the three-parameter size distribution (i.e., $\varepsilon_0,\,\varepsilon_1$ and $\amax$) in time, using a semi-analytic description of dust coagulation. The prescriptions given in this paper represent the first iteration of our new \tpop{} model that is derived and calibrated for vertically integrated disk models. Therefore, all calculations include gas and dust column densities instead of volume densities.

\subsection[Particle growth]{Particle growth (evolution of $\amax$)}
We model particle growth within the monodisperse approximation (\cref{eq:monogr}). The growth limits are realized by comparing a given fragmentation velocity $v_\text{frag}$ with the velocities between large grains \new{$\Delta v_\mathrm{max}$}, given by turbulence, differential settling and drift, and Brownian motion. 
For this, we modify our growth rate by a sigmoid-like function, which leads to growth for $\Delta v_\mathrm{max} < v_\text{frag}$, and decay for $\Delta v_\mathrm{max} > v_\text{frag}$, resulting in 
\new{\begin{equation}\label{eq:growthrate}
    \dot{a}_\mathrm{max} = \frac{\Sigma_1 \Delta v_\mathrm{max}}{\rho_\mathrm{m} \sqrt{2\pi} H_1}  \left(\frac{\left(\frac{v_\mathrm{frag}}{\Delta v_\mathrm{max}}\right)^s -1}{1 + \left(\frac{v_\mathrm{frag}}{\Delta v_\mathrm{max}}\right)^s}\right)\, ,
\end{equation}}\noindent
with $s$ being a parameter controlling the steepness of the transition from growth to fragmentation, $H_1$ being the scale height of large dust grains. \new{The relative grain velocity $\Delta v_\mathrm{max}$ is defined between grains of size $\amax$ and $f_{\Delta v} \amax$, where $f_{\Delta v}\in\left(0,1\right)$ is a model parameter. 
Determining \new{$f_{\Delta v}$}} and $s$ is the main task during the calibration of our model with respect to the full coagulation code \dpy{}\rev{\,(see \cref{sec:GrowthCalibration} and the red numbers in \cref{tab:1Dparams})}.

% The time evolution of the maximum dust size in our model is then realized via an explicit Euler step
% \begin{equation}\label{eq:growth}
%     a_{\mathrm{max}}^{(n+1)} = a_{\mathrm{max}}^{(n)} + \dot{a}\,\mathrm{d}t.
% \end{equation}

\subsection[Fragmentation and sweep-up]{Fragmentation and sweep-up (evolution of $\Sigma_0$ and $\Sigma_1$)}

In our model, two effects account for the evolution of the three-parameter size distribution and the interaction between the populations; fragmentation transfers mass from the large population to the small population, while collisions between larger and smaller particles lead to sweep-up, and thus mass transfer from the small bin to the large bin. Erosion of large particles due to collisions with small grains is not accounted for in our model. We describe the sweep-up process via the collision rates between large and small particles
\begin{equation}
    \dot{\rho}_{\text{d}, 0\rightarrow 1} = \frac{\rhog^2 \varepsilon_0 \varepsilon_1}{m_0 m_1}\sigma_{01} \Delta v_{01} m_0\, ,
\end{equation}
where $m_0$ and $m_1$ are the representative particle masses, $\Delta v_{01}$ is the representative relative velocity between the large and small particles, and $\sigma_{01}$ is the representative collision cross section.
These quantities are derived from the respective population's mass-averaged particle sizes (see \cref{eq:amean}).

Fragmentation predominantly occurs in collisions between two large grains. The corresponding transfer rate is thus determined by the collision rates between large grains
\begin{equation}
    \dot{\rho}_{\text{d}, 1\rightarrow 0} = \frac{\rhog^2 \varepsilon_1^2}{m_1^2}\sigma_{11} \Delta v_{11} m_1 \mathcal{F}\, .
\end{equation}
Here $\mathcal{F}$ represents a function that regulates the relative effectiveness of sweep-up and fragmentation. It is a function of the grain size, the desired power-law exponent of the size distribution, and the relative velocities $\Delta v_{11}$, \new{defined for collisions between particles of size $a_1$ and $f_{\Delta v}a_1$}. 
As we cannot model the microphysics of collisions between grains in our simplified framework for dust evolution, we base the functional form of $\mathcal{F}$ on the well-understood results of full coagulation models, which treat the evolution of the size distribution as the result of collisions between grains of all present sizes.

For this first version of our three-parameter model, we are only considering vertically integrated disk models, which means our mass transfer rates are given by 
\begin{align}
    \dot{\Sigma}_{0\rightarrow 1} &= \frac{\Sigma_0\Sigma_1 \sigma_{01}\Delta v_{01}}{ m_1\sqrt{2\pi (H_0^2+ H_1^2)}} \label{eq:rate01} \\
    \dot{\Sigma}_{1\rightarrow 0} &= \frac{\Sigma_1^2 \sigma_{11}\Delta v_{11}}{ m_1\sqrt{4\pi H_1^2}}\tilde{\mathcal{F}} \label{eq:rate10}\, ,
\end{align}
(see \cref{app:VertInt}).
In order to determine \rev{the vertically integrated version of} $\mathcal{F}$, named $\tilde{\mathcal{F}}$, we consider the steady state between fragmentation and sweep-up. In this equilibrium, a steady size distribution would be reached. 
Given $\dot{\Sigma}_{\text{d}, 0\rightarrow 1}=\dot{\Sigma}_{\text{d}, 1\rightarrow 0}$, we arrive at 
\begin{equation}
       \tilde{\mathcal{F}} =  \sqrt{\frac{2H_1^2}{H_0^2+H_1^2}}\frac{\sigma_{01}}{\sigma_{11}}\frac{\Delta v_{01}}{\Delta v_{11}}\left(\frac{\amax}{\aint}\right)^{- (q+4)}\, ,
\end{equation}
where $q$ is the desired power-law exponent of the distribution which will be reached on the dominating collisional timescale. 

Mass redistribution due to sweep-up and fragmentation is realized by defining the source terms of both populations in a total-mass-conserving manner as
\begin{align}
\begin{split}
        \dot{\Sigma}_{0} &= \dot{\Sigma}_{1\rightarrow 0} - \dot{\Sigma}_{0\rightarrow 1} \label{eq:redistr}\\
    \dot{\Sigma}_{1} &= - \dot{\Sigma}_{0}\, .
\end{split}
\end{align}

\rev{
\subsubsection*{Determining the size distribution power-law exponent $q$}
We have summed up the typical particle size distribution exponents in \cref{sec:dustcoag}, which were determined by \cite{Birnstiel2011} for distributions in coagulation-fragmentation equilibrium. These are given by 
\begin{align}
    q_\mathrm{frag} &= 
    \begin{cases}
         q_\mathrm{turb.1} = -3.75  &\text{small particle turb.\ regime} \nonumber \\
        q_\mathrm{turb.2}   = -3.5 &\text{fully intermediate turb.\ regime} \nonumber \\
        q_\text{drift-frag} = -3.75  &\text{drift-dominated regime} \nonumber \\
    \end{cases} \\
    q_\mathrm{sweep} &= -3.0  \quad \text{not in equilibrium \citep[see][]{Birnstiel2023}} \nonumber
\end{align}
Our task is now to find a way to smoothly switch between these regimes in our three-parameter coagulation model depending on which regime is prevailing under the given conditions.

Firstly, we can determine whether to apply the equilibrium size distributions $q_\mathrm{frag}$ or whether the dust has not yet reached coagulation-fragmentation equilibrium and collisions lead predominantly to sticking which results in $q_\mathrm{sweep}$. For this, we define a transition function
\begin{align}
\begin{split}
        p_{\text{frag}} &\coloneqq 
    \begin{cases}
    \rightarrow 1 \,\, \text{for} \,\, \Delta v_\mathrm{max}> v_{\text{frag}} \\ \rightarrow 0 \,\, \text{for} \,\, \Delta v_\mathrm{max}< v_{\text{frag}}
    \end{cases}
    \\
    p_{\text{sweep}} &= 1-p_{\text{frag}} \\
    \Rightarrow  q &=  q_{\text{frag}}\cdot  p_{\text{frag}} + q_{\text{sweep}}\cdot  p_{\text{sweep}} \, .
\end{split}
\label{eq:fragstick}
\end{align}

We now have to determine the equilibrium size distribution exponent $p_\mathrm{frag}$.
For this, we can again define transition functions. We determine whether the small particle turbulence regime (turb.1) or the fully intermediate regime (turb.2) dominates the relative turbulent velocities
\begin{align}
\begin{split}
        p_{\text{turb.1}} &\coloneqq 
    \begin{cases}
    \rightarrow 1 \,\, \text{for} \,\, \Delta v_\text{turb.1}>\Delta v_\text{turb.2} \\ \rightarrow 0 \,\, \text{for} \,\, \Delta v_\text{turb.1}< \Delta v_\text{turb.2}
    \end{cases}
    \\
    p_\text{turb.2} &= 1-p_\text{turb.1} \\
    \Rightarrow  q_\text{turb-frag} &=  q_\text{turb.1}\cdot  p_\text{turb.1} + q_\text{turb.2}\cdot  p_\text{turb.2} \, .
\end{split}
\label{eq:turb-frag}
\end{align}

Lastly, we must determine whether we are in the turbulence-dominated regime or in the drift-dominated regime. Similar to before, we define
\begin{align}
\begin{split}
        p_{\text{drift}} &\coloneqq
    \begin{cases}
    \rightarrow 1 \,\, \text{for} \,\, \Delta v_\text{drift}>\Delta v_\text{turb} \\ \rightarrow 0 \,\, \text{for} \,\, \Delta v_\text{drift}< \Delta v_\text{turb}
    \end{cases}
    \\
    p_\text{turb} &= 1-p_\text{drift} \\
    \Rightarrow  q_\mathrm{frag} &=  q_\text{drift-frag}\cdot  p_\text{drift} + q_\text{turb-frag}\cdot  p_\text{turb} \, ,
\end{split}
\label{eq:drift-frag}
\end{align}
where $q_\text{turb-frag}$ comes from \cref{eq:turb-frag}. With this, we have everything we need to calculate $q$ from \cref{eq:fragstick} and we can determine the mass exchange rates from \cref{eq:redistr}.
The exact form of the transition functions is not of great importance as long as the transition is sufficiently fast but still smooth enough to not cause issues during numerical integration. The choices that worked best in our subsequent tests are listed in \cref{sec:transfunc}.
}

\subsection{Passive dust fluids (in the \pluto{} code)}
\label{sec:dusttransport}
We use the \pluto{}\footnote{\url{http://plutocode.ph.unito.it/}} code to solve the equations of hydrodynamics in our calibration and test simulations with \tpop{}.
The Euler equations, solved by \pluto{}, read
\begin{align}
\frac{\partial \rhog}{\partial t}+\vec{\nabla}\cdot (\rhog\vec{v})&=0 \\
\frac{\partial \rhog \vec{v}}{\partial t}+\vec{\nabla}\cdot (\rhog\, \vec{v}\otimes\vec{v})&=-\vec{\nabla}P - \rhog \vec{\nabla}\Phi\, \rev{+\vec{\nabla}\cdot \mathbb{T}} \, ,
%&=\rev{+\nu_1\left(\Vec{\nabla}\Vec{v} + \left(\Vec{\nabla}\Vec{v}\right)^\mathrm{T}\right) + \left(\nu_2-\frac{2}{3}\nu_1\right)(\Vec{\nabla}\cdot \Vec{v})\mathbb{I}}\, , 
\end{align}
where $\vec{v}$ is the gas velocity vector, $\Phi$ is the gravitational potential\rev{, and $\mathbb{T}$ is the viscous stress tensor.}
The ideal equation of state is used as a closure relation
\begin{equation}
   P=\frac{k_\text{B} T}{\mu m_{\text{p}}}\rhog\, .
\end{equation}

The \pluto{} code allows for the treatment of passive tracer fluids, which are simply advected with the gas following
\begin{equation}
    \der{(\rhog \varepsilon)}{t} + \vec{\nabla} \cdot (\varepsilon \,  \rhog \, \vg) = 0\, .
\end{equation}
In this work, we consider vertically integrated protoplanetary disks and the advected quantities in our \tpop{} model are thus the local dust-to-gas ratios of our two dust populations $\varepsilon_0 =\Sigma_0/\Sigmag$ and $\varepsilon_1 =\Sigma_1/\Sigmag$. The maximum particle size is defined as a tracer of the large dust population, meaning our third tracer fluid is $\amax\varepsilon_1$. 

The respective tracer fluxes are modified to simulate a dust fluid that is aerodynamically coupled to the gas and thus undergoes radial and azimuthal drift (in the terminal velocity approximation). 
To achieve this within \pluto{}'s tracer prescription, we add a flux component corresponding to the relative velocity between dust and gas \rev{
\begin{equation}
    \Vec{v}_{\Sigma_{0/1}} = \frac{\St_{\Sigma_{0/1}}}{\St_{\Sigma_{0/1}}^2+1}\frac{1}{\OmK \rhog} \Vec{\nabla} P\, ,
\end{equation}
where $\St_{\Sigma_{0/1}}$ is the mass-averaged Stokes number of the respective population.} The drift velocities are limited to a fraction of the soundspeed. The third tracer ($\amax \varepsilon_1$) is given the same drift velocity as the large dust population ($\varepsilon_1$).
The tracer fluxes are calculated with the upstream dust density and \rev{density-weighted maximum particle size} based on the drift velocities at the respective cell interface as
\begin{align}\label{eq:dr_flux}
\vec{F}^\mathrm{drift}_{\Sigma_0, \, i+\nicefrac{1}{2}} =&\,\Sigma_{0, \, i}\, \max(0, v_{\Sigma_0, \, i+\nicefrac{1}{2}}) + \Sigma_{0, \, i+1}\, \min(v_{\Sigma_0, \, i+\nicefrac{1}{2}}, 0) \\
\vec{F}^\mathrm{drift}_{\Sigma_1, \, i+\nicefrac{1}{2}} =&\,\Sigma_{1, \, i}\, \max(0, v_{\Sigma_1, \, i+\nicefrac{1}{2}}) + \Sigma_{1, \, i+1}\, \min(v_{\Sigma_1, \, i+\nicefrac{1}{2}}, 0) \\
\vec{F}^\mathrm{drift}_{\amax\Sigma_1, \, i+\nicefrac{1}{2}} =&\,a_{\mathrm{max},\,i}\Sigma_{1, \, i}\, \max(0, v_{\Sigma_1, \, i+\nicefrac{1}{2}}) \nonumber \\
&+ a_{\mathrm{max},\,i+1}\Sigma_{1, \, i+1}\, \min(v_{\Sigma_1, \, i+\nicefrac{1}{2}}, 0)\, .
\end{align}

% \cite{Youdin2002}, derived an analytical prescription for the time evolution of the dust column density in a protoplanetary disk with the method of characteristics. Given that the drift velocity can be described as a powerlaw $\vdr=v_0 (R/R_0)^d$, they find 
% \begin{align*}
% \Sigma(R,t) &= \Sigma(r_0,0)\frac{\vdr(R_0) R_0}{\vdr(R)R} \\  
% R_0 &= R\left(1-(1-d)\frac{\vdr(R)t}{R}\right)^\frac{1}{1-d}.
% \end{align*}
% We use this analytical result as a test case for our passive dust fluid in \pluto{}. 

Dust diffusion is implemented in a flux-limited manner, where the transport velocity is limited to the turbulent gas velocity $v_\text{max}=\frac{\sqrt{\delta}c_\mathrm{s}}{1+\St^2}\, $. The diffusion fluxes are given by  
\new{
\begin{align}
\vec{F}^{\text{diff}}_{\Sigma_0, i+\nicefrac{1}{2}} &= -\lambda_{\mathrm{lim}} \frac{\delta c_\mathrm{s} H}{1+\St_0^2} \Sigma_{\mathrm{g}, i+\nicefrac{1}{2}} \vec{\nabla}\varepsilon_0 \\
\vec{F}^{\text{diff}}_{\Sigma_1, i+\nicefrac{1}{2}} &= -\lambda_{\mathrm{lim}} \frac{\delta c_\mathrm{s} H}{1+\St_1^2}  \Sigma_{\mathrm{g}, i+\nicefrac{1}{2}} \vec{\nabla}\varepsilon_1 \\
    \vec{F}^{\text{diff}}_{\amax \Sigma_1,i+\nicefrac{1}{2}} &= -\lambda_{\mathrm{lim}} \frac{\delta c_\mathrm{s} H}{1+\St_1^2} \Sigma_{\mathrm{g}, i+\nicefrac{1}{2}} \vec{\nabla}(\amax\varepsilon_1) \, , 
    \label{eq:diff_flux}
\end{align}
}\noindent
where $\lambda_\mathrm{lim}$ is the respective flux limiter function \rev{(see \cref{sec:fluxlimiter})}. \new{All quantities are interpolated to the cell interface}.
The drift and diffusion fluxes are added to the advective flux of the dust fluids stemming from the gas motion. We have used this approach in \cite{Pfeil2023}, where we also presented some simple test cases of the method. 
\rev{Note that this approach makes use of the terminal velocity approximation and is thus strictly speaking only valid for $\St \ll 1$. The \tpop{} model itself, being a local model, is not bound to this form of dust advection scheme and could be implemented in any multi-fluid-capable hydrodynamics code.}

\subsection{Complete right-hand side of the conservation equations (in the \pluto{} code)}
All modifications made for our three-parameter dust evolution model can be applied within the framework of \pluto{} and without changing the underlying reconstruct-solve-average scheme of the code. 

Source terms are added to the right-hand side of the conservative hydrodynamics equations describing the evolution of the three-parameter dust size distribution in the framework of \pluto{}. For each dimension and evolving variable, we have the right-hand side of the conservation equation
\begin{equation}
    \mathcal{R}_i = -\frac{\Delta t}{\Delta \mathcal{V}_i}\left[(\mathcal{A}F)_{i+\nicefrac{1}{2}}-(\mathcal{A}F)_{i-\nicefrac{1}{2}}\right] + \Delta t \mathcal{S}_i\, ,
\end{equation}
where $\Delta t$ is the timestep, $\Delta \mathcal{V}$, is the cell volume, $\mathcal{A}$ the respective cell surface, $F$ is the flux through the interfaces, determined from advection with the gas plus relative terminal velocity and diffusion fluxes, and $\mathcal{S}_i$ is the source term, given by fragmentation, sweep-up, and growth.

For the two dust fluids, the fluxes are calculated via \cref{eq:dr_flux} to \cref{eq:diff_flux}, and the source term of the respective dust fluid is calculated from \cref{eq:redistr}
\begin{align}
    F_{\Sigma_{0/1},\, i\pm\nicefrac{1}{2}} &= F^\mathrm{diff}_{\Sigma_{0/1}, i\pm\nicefrac{1}{2}} + F^\mathrm{drift}_{\Sigma_{0/1}, i\pm\nicefrac{1}{2}} + F^\mathrm{adv}_{\Sigma_{0/1}, i\pm\nicefrac{1}{2}}\\
    \mathcal{S}_{\Sigma_{0/1},\,i} &= \dot{\Sigma}_{0/1,\, i}\, ,
\end{align}
where $F^\mathrm{adv}_{\Sigma_{0/1}, i\pm\nicefrac{1}{2}}$ is the flux component that is calculated by \pluto{} for the passive advection of the tracers.
The maximum particle size is advected together with a large population as a tracer. The fluxes and growth rate are
\begin{align}
F_{\amax\Sigma_1, i\pm\nicefrac{1}{2}} &=  F^\mathrm{diff}_{\amax\Sigma_{1}, i\pm\nicefrac{1}{2}} + F^\mathrm{drift}_{\amax\Sigma_{1}, i\pm\nicefrac{1}{2}} + F^\mathrm{adv}_{\amax\Sigma_{1}, i\pm\nicefrac{1}{2}}\\
    \mathcal{S}_{\amax\Sigma_1,\, i} &= a_{\mathrm{max},\,i}\dot{\Sigma}_{1,\, i}+\Sigma_{1,\, i}\dot{a}_{\mathrm{max},\, i}\, .
\end{align}
\rev{Here, $F^\mathrm{adv}_{\amax\Sigma_{1}, i\pm\nicefrac{1}{2}}$ is the flux component that is calculated by \pluto{} for the passive advection of the tracer and $\dot{a}_\mathrm{max}$ is the particle growth rate, given by \cref{eq:growthrate}.}
The evolution equations are evolved in time with \pluto{}'s standard third-order Runge-Kutta routine. The timestep is limited by the CFL condition, which we don't have to adjust since the dust drift velocities are limited to a fraction of the speed of sound and the typical growth timescale is much longer than an orbital timescale.

\section{Calibration}
\label{sec:calibration}
Our model has several free parameters, which have to be calibrated in comparison to full dust coagulation simulations.
For the growth rate of the dust (\cref{eq:growthrate}), the main parameters are $s$ (determining the steepness of the transition from growth to fragmentation), and the parameter $f_{\Delta v}$, which determines the relative velocities $\Delta v_{11}$ \new{and $\Delta v_\mathrm{max}$} through the size ratio between the grains.

We ran a series of one-dimensional simulations for calibration of the model against the full-coagulation code \dpy{}. We set up a 1D disk model similar to the standard \dpy{} model. We excluded viscous evolution since we are only interested in comparing the dust evolution in both models for the time being. 
The radial disk structure follows
\begin{align}
\Sigma_\mathrm{g}(R)&=(2+\beta_{\Sigma})\frac{M_{\text{disk}}}{2\pi R_\mathrm{c}^2} \left(\frac{R}{R_\mathrm{c}}\right)^{\beta_{\Sigma}}\exp{\left[-\left(\frac{R}{R_\mathrm{c}}\right)^{2+\beta_{\Sigma}}\right]} \nonumber \\
&\coloneqq \Sigma_\mathrm{g,0}
\left(\frac{R}{R_0}\right)^{\beta_{\Sigma}}\exp{\left[-\left(\frac{R}{R_\mathrm{c}}\right)^{2+\beta_{\Sigma}}\right]}\, ,
\end{align}
where we employed the code units $\Sigma_\mathrm{g,0}=\SI{733.28}{\gram \per \square \centi \meter}$ (derived from \rev{a} \dpy{} setup with $M_\mathrm{disk}=\SI{0.05}{\Msol}$), $\beta_\Sigma=-0.85$, $\beta_T=-0.5$, and $R_0=\SI{1}{\AU}$. In \pluto{}, the radial temperature structure is expressed in terms of the speed of sound, which is given by
\begin{align}
    c_\mathrm{s}^2 &=T_0\frac{k_\text{B}}{\mu m_\text{p}} = \left(\frac{0.05 L_*}{4\pi R^2 \sigma_\text{SB}}\right)^\frac{1}{4}\frac{k_\text{B}}{\mu m_\text{p}} \nonumber  \\
    &\coloneqq v_0^2 \left(\frac{H_0}{R_0}\right)^2 \left(\frac{R}{R_0}\right)^{\beta_T}\, ,
\end{align}
with $v_0=\sqrt{GM_*/R_0}\, $. In our simulations, this was parameterized by the disk aspect ratio $\nicefrac{H_0}{R_0}$ at reference radius $R_0$.

The simulations were initialized with a total dust-to-gas ratio of 1\%, and an initial maximum particle size of \SI{1}{\micro \meter}. Particles larger than the initial drift limit were excluded from the initial dust profile to avoid an inwards drifting wave of large dust at the beginning of the simulation. The initial dust size distribution in both \dpy{} and \tpop{} follows a power law with $q=-3.5\, $.
The stellar and disk structure parameters are shown in \cref{tab:StarAndDiskParameters} and the parameters for the different calibration runs are shown in \cref{tab:1Dparams}.

\begin{table}[t]
\caption{Stellar parameters and disk parameters for the simulations presented in this work.}
    \centering
    \begin{tabular*}{\columnwidth}{@{\extracolsep{\fill}}ccccccc}
    \toprule
    $M_*$ & $R_*$ & $T_*$ & $M_\mathrm{disk}$ & $R_\mathrm{c}$  & $\beta_\Sigma$ & $\beta_T$ \\
    $[M_\mathrm{\odot}]$ & $[R_\mathrm{\odot}]$ & $\mathrm{[K]}$ & $[M_*]$ & $\mathrm{[au]}$  &  &  \\
    \midrule 
    \midrule
      1.0   & 3.096 & 4397 & 0.05 & 60.0 & -0.85 & -0.5 \\
      0.9   & 2.906 & 4315 & "    & "    & "     & "    \\
      0.7   & 2.678 & 4111 & "    & "    & "     & "    \\
      0.5   & 2.458 & 3849 & "    & "    & "     & "    \\
      0.3   & 2.215 & 3460 & "    & "    & "     & "    \\
      0.1   & 1.013 & 2925 & "    & "    & "     & "    \\
      \bottomrule
    \end{tabular*}
    \tablefoot{Calibration runs are performed for the one-solar-mass case.}
    \label{tab:StarAndDiskParameters}
\end{table}
\begin{table}[t]
\caption{Dust properties for the calibration runs (\cref{sec:GrowthCalibration}, \cref{sec:TransportCalibration}) and the test simulations with different stellar masses \cref{sec:StellarMasses}.}
    \centering
    \begin{tabular*}{\columnwidth}{@{\extracolsep{\fill}}lr}
    \toprule 
    Dust Property & Value \\
    \midrule
    \midrule
        Dust-to-gas ratio & 0.01 \\
        Initial \rev{maximum} dust size & \SI{e-4}{\centi \meter} \\
        Minimum dust size & \SI{e-5}{\centi \meter} \\
        Dust material density & \SI{1.67}{\gram \per \cubic \centi \meter} \\
        Dust fragmentation velocity & \SI{1000}{\centi \meter \per \second} \\
        \bottomrule
    \end{tabular*}
    \label{tab:dustproperties}
\end{table}

\begin{table}[t]
\centering
\caption{Parameters of the one-dimensional simulations performed to calibrate the three-parameter model to the \dpy{} simulations.}
\begin{tabular*}{\columnwidth}{@{\extracolsep{\fill}}ccccc}
\toprule
\multicolumn{3}{c}{Model Parameters} & \multicolumn{2}{c}{Physical Parameters} \\
$f_{\Delta v}$ & s & $f_\text{drift}$ & $\alpha$ (coagulation) & $\delta$ (diffusion) \\
\midrule
\midrule
\multicolumn{5}{c}{Upper row of \cref{fig:OneAmax}} \\
0.2  & 3 & \text{no transport} & $10^{-3}$ & \text{no transport}  \\
0.3  & " & " & " & "  \\
{\color{Maroon} \bfseries 0.4}  & " & " & " & "  \\
0.5  & " & " & " & "  \\
0.6  & " & " & " & "  \\
\midrule
\multicolumn{5}{c}{\cref{fig:SfacStudy}} \\
0.4& 2 & \text{no transport} & $10^{-3}$ & \text{no transport}  \\
"  & {\color{Maroon} \bfseries 3} & " & " & "  \\
"  & 4 & " & " & "  \\
"  & 5 & " & " & "  \\
"  & 6 & " & " & "  \\
\midrule
\midrule
\multicolumn{5}{c}{\cref{fig:FudgeDiff}} \\
0.4& 3 & 0.6 & $10^{-3}$ & $10^{-3}$  \\
"  & " & 0.7 & " & "  \\
"  & " & {\color{Maroon} \bfseries 0.8} & " & "  \\
"  & " & 0.9 & " & "  \\
"  & " & 1.0 & " & "  \\
\midrule
\multicolumn{5}{c}{\cref{fig:FudgeNoDiff}} \\
0.4& 3 & 0.6 & $10^{-3}$ & 0.0  \\
"  & " & 0.7 & " & "  \\
"  & " & {\color{Maroon} \bfseries 0.8} & " & "  \\
"  & " & 0.9 & " & "  \\
"  & " & 1.0 & " & "  \\

\bottomrule
\end{tabular*}
\tablefoot{ Each simulation in one block is compared to the same \dpy{} simulation with identical $\alpha$ and $\delta$. \rev{The bold red numbers are our final calibrated values used in the following test simulations.}}
\label{tab:1Dparams}
\end{table}

\subsection{Calibrating the growth rate}
\label{sec:GrowthCalibration}

\begin{figure*}[t]
    \includegraphics[width=\textwidth]{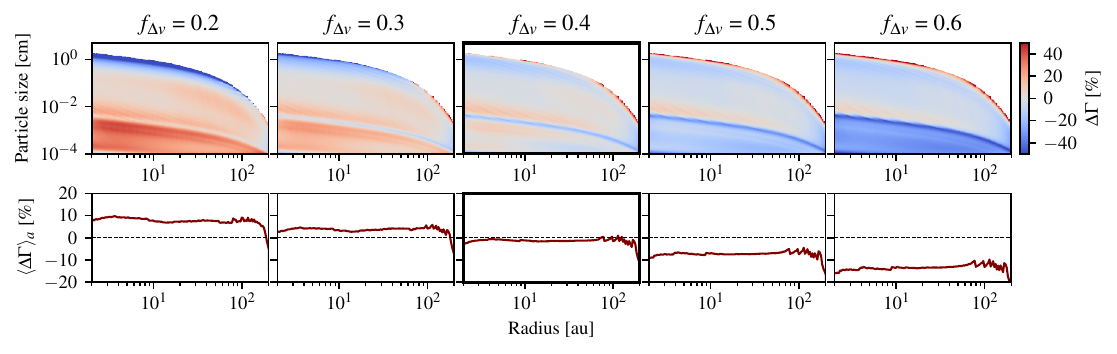}
    \caption[Dust growth maps parameter studies.]{Parameter study for the factor $f_{\Delta v}$, which is the most important parameter for the growth rate calibration of our model. For this, we calculated the deviation between the mass-averaged particle size's growth rate in \dpy{} and \tpop{}. In the top row, we show models for different global values of $f_{\Delta v}$. It can be seen that $f_{\Delta v}=0.4$ seems to fit the dust growth rate best.}
    \label{fig:OneAmax}
\end{figure*}
\begin{figure*}[t]
    \centering
    \includegraphics[width=\textwidth]{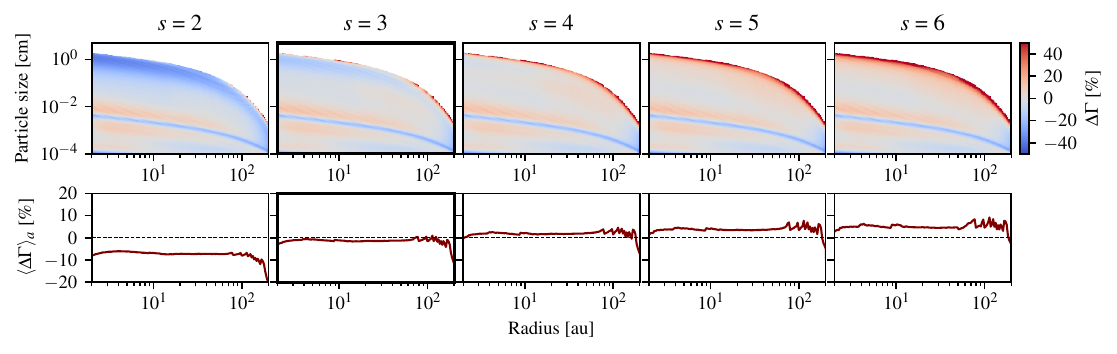}
    \caption[Dust growth maps for different $s$-factors.]{Parameter study for the factor $s$, which determines the steepness of the transition from sweep-up growth to coagulation-fragmentation equilibrium in \cref{eq:growthrate}. Heat maps show the deviation of the growth rate in \tpop{} from the growth rate in \dpy{}. The lower row shows the average over the size dimension.
    The influence of a variation in $s$ can be seen around the fragmentation limit, where a larger $s$ corresponds to a steeper transition (fast growth close to the fragmentation limit), while a small $s$ corresponds to a slower transition where growth rates decrease earlier and more slowly. The case of $s=3$ has the overall best fit with the \dpy{} model.}
    \label{fig:SfacStudy}
\end{figure*}

Determining the appropriate relative grain velocities to reproduce the full coagulation model with our simplified prescription is the topic of this section.

The particle growth rate in our model can be adjusted by varying the relative velocity \new{$\Delta v_\mathrm{max}$} in \cref{eq:growthrate}. This is controlled over the parameter $f_{\Delta v}\in\left(0,1\right)$, which sets the size ratio between the colliding particles \new{$\amax$ and $f_{\Delta v}\amax$, ($\Delta v_\mathrm{max}$) as well as $a_1$ and $f_{\Delta v}a_1$, ($\Delta v_{11}$)}. \rev{We have thus determined which particle size ratio best reproduces the relative velocity that determines the overall growth rate}.
Furthermore, the parameter $s$ in \cref{eq:growthrate} can be used to adjust the growth rate around the transition from growth to fragmentation, where a small $s$ corresponds to a reduced growth rate close to the fragmentation limit, and a large $s$ corresponds to a steep transition from particle growth to fragmentation-coagulation equilibrium.

To characterize the effects in comparison with the full coagulation model \dpy{}, we measured the rate of change of the mass-averaged particle size in \tpop{} and \dpy{} and determined the respective deviations for the given parameters. We used the mass-averaged size and not the maximum particle size because it is also a measure of the shape of the size distribution itself and not just the upper cutoff.
To calibrate the growth rate we ran \dpy{} and \tpop{} models without transport (all fluxes are zero). We were, however, still considering the relative drift and sedimentation velocities in the calculation of \new{$\Delta v_{01},\,\Delta v_{11}$ and $\Delta v_\mathrm{max}$}. 
We set up the disk models for a solar-mass pre-main-sequence star with a \SI{0.05}{\Msol} gas disk and a dust-to-gas ratio of \SI{1}{\percent} (see \cref{tab:StarAndDiskParameters} and \cref{tab:dustproperties} for details). 

First, we ran a parameter study for the factor $f_{\Delta v}$. \revI{This parameter can be physically interpreted as the particle size ratio in the collisions that dominate the growth process. Since we know that growth is dominated by particle collisions where the size difference between projectile and target is not too large, we chose values between \new{0.2 and 0.6}.} As can be seen in \cref{fig:OneAmax}, a factor of \new{$f_{\Delta v}=0.4$} leads to the overall best agreement between the particle growth rate in \tpop{} and \dpy{}. 

The second parameter setting the growth rate is $s\, $, which determines the steepness of the transition from growth to fragmentation in \cref{eq:growthrate}. \revI{We know that this transition is usually not leading to an instantaneous cutoff of the growth rate. In reality, the particles have collision velocities following a Maxwell-Boltzmann distribution \citep{Stammler2022}. Very large values of $s$ would thus lead to a time evolution that does not agree well with \dpy{}.}
We ran models with values from 2 to 6, which we compared in \cref{fig:SfacStudy}. As can be seen, a value of $s=3$ leads to the best fit between the growth rates close to the fragmentation limit.

\subsection{Calibrating the dust transport}
\label{sec:TransportCalibration}
\begin{figure*}[t]
    \includegraphics[width=\textwidth]{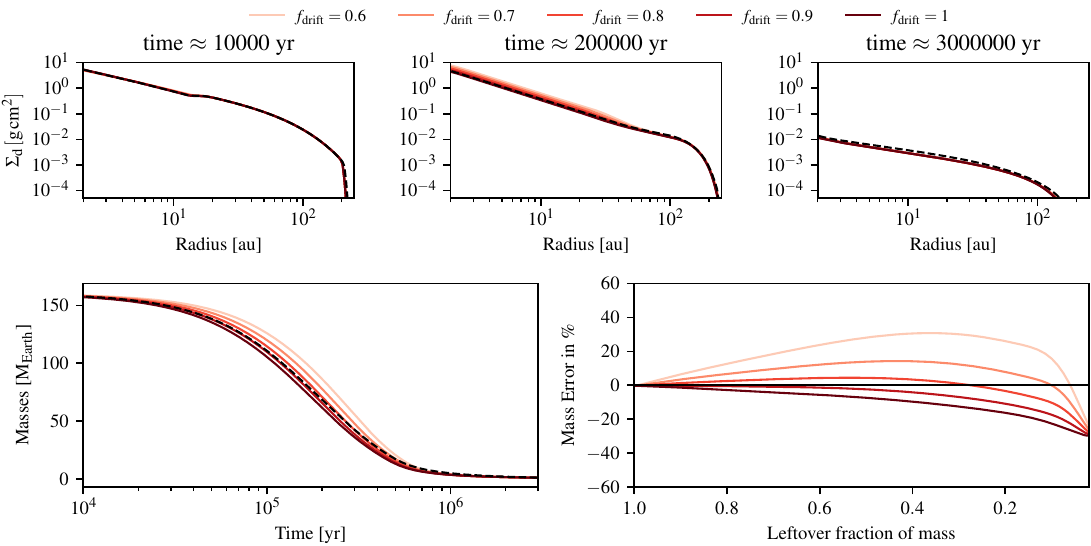}
    \caption[Drift velocity factor parameter study.]{Comparison between \dpy{} and our model in a setup with dust diffusion ($\delta=10^{-3}$) and with different drift calibration factors $f_\mathrm{drift}$. \revI{Solid lines show the results of our \tpop{} calibration runs and dashed lines show the respective \dpy{} simulation, to which we calibrate our model.} The upper row shows a timeseries of the dust column density evolution in three snapshots. In the lower row, we show the mass evolution and the errors with respect to the full coagulation model \dpy{}. For a factor of \new{$f_\mathrm{drift}=0.8$}, the mass evolution of the full coagulation model is well reproduced by our three-parameter model.}
    \label{fig:FudgeDiff}
\end{figure*}

Dust transport is realized by modifying the dust tracer flux (see \cref{sec:dusttransport}). For this, the mass-averaged particle size of each dust population is calculated to determine the upstream flux through the respective cell interface. 
\new{This means that the steepness of the size distribution, which determines the amount of mass in each bin, also partially determines the radial mass flux. As discussed in the introduction, the power law's shape in fragmentation-coagulation equilibrium is well known. In the drift limit, however, it can not be easily deduced from analytical arguments. Numerical dust coagulation simulations with \dpy{} show that the power law is generally more top-heavy in the drift limit/sweep-up phase and probably best approximated by a value between -2.5 and -3. We have therefore conducted \tpop{} simulations covering this parameter range to determine what value best reproduces the late-stage mass evolution of the disk. The mass evolution is generally well reproduced for $q_\mathrm{sweep} = -3.0$, which is why this value is used for all following simulations.}

In comparison with a more realistic size distribution, the power-law prescription lacks the gradual decrease in mass, shortly before the maximum particle size (visible in the right panels of \cref{fig:fiducial}). 
This means that, even if our power-law size distribution reproduces the real size distribution very well, the mass-averaged sizes of both models will not be identical.
For this reason, our dust fluxes will also be slightly different from the total flux in \dpy{}. This is generally not a significant effect. We nonetheless accounted for it by using a calibration factor $f_\mathrm{drift}$ that is multiplied to the calculated mass-averaged particle sizes before the calculation of the dust flux, similar to the fudge factor in the \tpoppy{} model \citep{Birnstiel2012}. 
For this, we ran models for different values of this calibration factor, including dust transport. We neglected any gas transport in these simulations and focused on the dust fluxes, where we conducted one set of simulations with a dust diffusivity of $\delta=10^{-3}$ and one with $\delta=0$ (no diffusion). 
Except for the inclusion of dust transport, the simulations are identical to the setups used in the growth rate calibration. We applied the best-fitting parameters from the growth rate calibration.
We use 150 \rev{log-spaced} radial grid cells to resolve our simulation domain, which was defined between \SI{2}{\AU} and \SI{250}{\AU}. This is the same resolution as used in the \dpy{} runs.
For comparison, we present a time series of the dust column densities in these simulations runs for $f_\mathrm{drift}=$\SIlist{0.6;0.7;0.8;0.9;1.0}{} in the upper row of \cref{fig:FudgeDiff} (\cref{fig:FudgeNoDiff} depicts the case without diffusion). The lower two panels depict the dust mass evolution (lower left) and the respective deviations from the full coagulation model \dpy{} (lower right).
As can be seen, the choice of $f_\mathrm{drift}$ determines the mass flux throughout the disk's dust evolution. Smaller values of $f_\mathrm{drift}$ correspond to slower dust velocities and lower fluxes. All values result in deviations of $\lesssim$\SI{40}{\percent} from the full coagulation model at all times. The value of \new{$f_\mathrm{drift}=0.8$}, however, shows the smallest mass error. For this value, the disk's dust mass only deviates by \SI{10}{\percent} from \dpy{} during most of the mass evolution. The absolute error only increases to $\lesssim\SI{30}{\percent}$ after \SI{90}{\percent} of the mass has already drifted out of the simulation domain.
The overall trend in the mass evolution with $f_\mathrm{drift}$ seems to be independent of the diffusion parameter, as can be seen by comparison of \cref{fig:FudgeDiff} and \cref{fig:FudgeNoDiff}.
We therefore chose a value of \new{$f_\mathrm{drift}=0.8$} for all following simulations.

\rev{
\subsection{Treatment of planetary gaps}
\begin{figure}[t]
    \centering
    \includegraphics[width=\columnwidth]{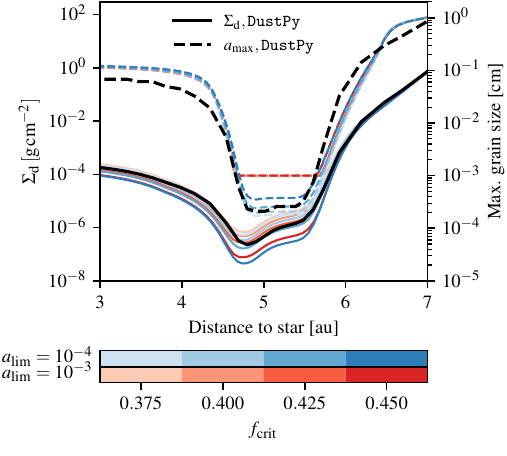}
    \caption[Calibration of the grain size reduction in planetary gaps.]{Test for the size reduction in planetary gaps. \revI{The dashed lines correspond to the maximum grain size of the respective model (see color bars for the model parameters), while the solid lines show the dust column density for each model. Thick \revII{black} lines correspond to the respective \dpy{} simulation that we want to reproduce.} The particle size is reduced on the dust depletion timescale when the large dust population makes up less than the critical fraction $f_\mathrm{crit}$ of the total dust density. We run these simulations to determine which value leads to the best agreement with the \dpy{} simulation.}
    \label{fig:TestShrink}
\end{figure}
\label{sec:GapCalib}
Since the particle size in \tpop{} can only change due to coagulation/fragmentation, we have to introduce an additional source term that takes into account the particle size reduction due to the depletion of large particles via transport. The classic example of such a case is a gap carved by a planet. In this case, dust diffusion into the gap from the outer disk, removal of large grains within the gap, and coagulation determine what particle sizes are present. \revI{Due to the pressure maximum at the gap's outer edge, larger grains can no longer drift into the co-orbital region of a planet. However, smaller particles are still diffused through the pressure maximum by turbulence and thus slightly increase the solid content of the mostly depleted gap.}
As the largest particles are removed from the gap, we reduce the maximum particle size on the respective dust depletion timescale in \tpop{}. For this, we set a lower limit for the fraction of large particles at which we begin the size reduction process. We define a hypothetical source term for $\varepsilon_1$, which would set a lower limit to $\varepsilon_1$ in terms of a critical fraction of the total dust density $\frac{\mathrm{d} \varepsilon_1}{\mathrm{d} t} = \frac{f_\mathrm{crit}(\varepsilon_0+\varepsilon_1)-\varepsilon_1}{\Delta t}$
where $\Delta t$ is the current simulation timestep. Since our goal is to reduce the maximum particle size on the depletion timescale, we set $\tau_\mathrm{depletion}=\nicefrac{\varepsilon_1}{\frac{\mathrm{d}\varepsilon_1}{\mathrm{d} t}}$ and define the respective size reduction rate
\begin{equation}
    \frac{\mathrm{d}\amax}{\mathrm{d}t}=\frac{\amax}{\tau_\mathrm{depletion}}\left(1-\frac{\amax}{a_\mathrm{lim}}{}\right)\, ,
    \label{eq:Shrink}
\end{equation}
where $a_\mathrm{lim}=\SI{1}{\micro \meter}$ is a minimum size that we define to limit the shrinking to a reasonable value. To retain a meaningful power-law exponent for the size distribution during the shrinking process, we set the mass change rate of the large population to
\begin{equation}
    \frac{\mathrm{d}\varepsilon_1}{\mathrm{d} t} = \frac{\partial\varepsilon_1}{\partial \amax}\frac{\mathrm{d}\amax}{\mathrm{d} t}\, .
    \label{eq:ShrinkMass}
\end{equation}
Here, $\nicefrac{\partial\varepsilon_1}{\partial \amax}$ follows analytically from \cref{eq:distr}. \Cref{eq:Shrink} and \cref{eq:ShrinkMass} are added as source terms to the respective right-hand side of the conservation equations if the condition for size reduction is fulfilled ($\varepsilon_1<f_\mathrm{crit}\varepsilon_\mathrm{tot}$).
In that way we achieve the following:
\begin{itemize}
    \item Whenever transport is largely dominating over coagulation, thus removing the large particles, we reduce the maximum particle size on the depletion timescale of the large grain population. 
    \item We thus set a lower limit for the power-law exponent and therefore retain a physically meaningful size distribution, even in planetary gaps.
\end{itemize}
Determining the critical value of $\varepsilon_1$ at which we begin the size reduction process is an experimental task. The size distribution within a planetary gap is typically not top-heavy due to the efficient removal of the largest grains \citep[e.g.,][]{Drazkowska2019}. The large mass bin should thus contain $\lesssim \SI{50}{\percent}$ of the total dust density in the gap.

We set up a one-dimensional simulation with a gap corresponding to a Jupiter-mass planet, following the description of \cite{Duffell2020} in \tpop{} and \dpy{}. The gap is not evolving in time, but pre-defined in the disk's gas structure and the initial dust density structure.
We ran simulations for several values of the critical mass fraction $f_\mathrm{crit}$ to determine which values give us the best fit between \tpop{} and \dpy{} in a planetary gap. For the limiting particle size, we set either \SI{1}{\micro \meter} or \SI{10}{\micro \meter}. The resulting dust column density profiles are shown in \cref{fig:TestShrink}.
As can be seen, we reached a reasonably good fit for a value of \new{$f_\mathrm{crit}=0.425$} and a limiting particle size of $a_\mathrm{lim}=\SI{1}{\micro \meter}$. We have therefore chosen these values for all following simulations. 
}

\section{Test Simulations}
\label{sec:tests}

Here, we present a series of test cases of our \tpop{} model for comparison with \dpy{}.  Our fiducial model is depicted in \cref{fig:fiducial}. We set up a smooth protoplanetary disk around a solar-mass star for this. The disk structure is identical to the simulations in the calibration runs.
The first two panels in each row show the snapshots of the dust size distributions in \dpy{} (top row) and \tpop{} after  \SI{20500}{\years} and \SI{3e6}{\years} of evolution. For this run, gas evolution has been turned off to get a more direct comparison between the dust evolution and transport models.
The shape of the size distribution is well reproduced in the drift-limited and fragmentation-limited cases.

\subsection{Different stellar masses}
\label{sec:StellarMasses}
\begin{figure*}[p]
    \includegraphics[width=\textwidth]{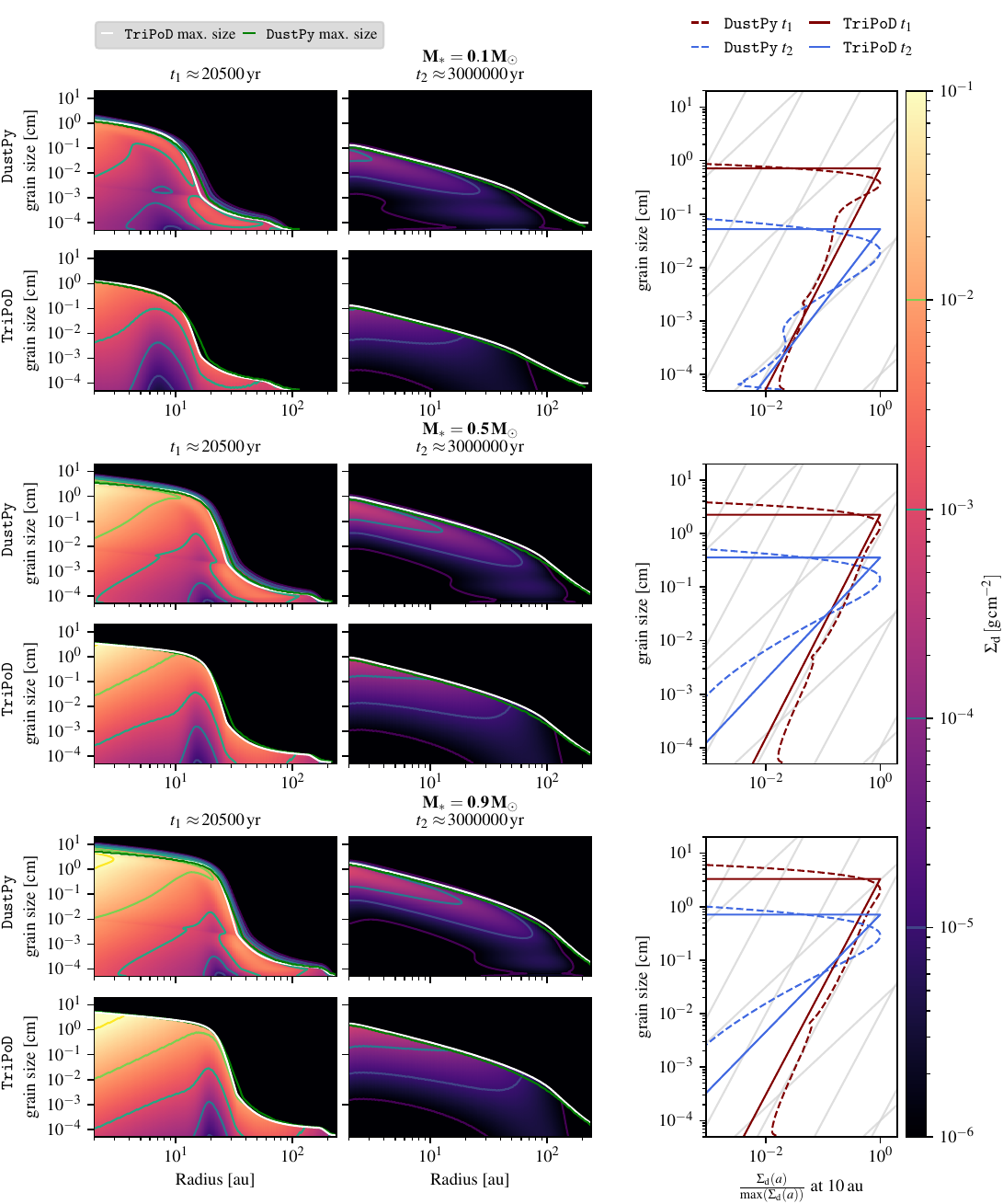}
    \caption[Dust size distributions for three different stellar masses with \dpy{} and \tpop{}.]{Comparison of the dust size distributions in three simulations with different stellar masses and different disk masses. The first row in each group depicts a \dpy{} simulation (full treatment of coagulation). The lower rows show the respective \tpop{} simulations with the \pluto{} code. Panels on the right depict the local dust size distributions at \SI{10}{\AU} at both snapshots. The gray lines in the right panel follow power laws with $q=-3.5$ and $q=-3.0$.}
    \label{fig:StellarMasses}
\end{figure*}

We tested the accuracy of our model in simulations of systems with different stellar masses and disk masses. For this, we took the stellar properties of stars with $M_*=\,$\SIlist{0.1;0.3;0.5;0.7;0.9}{\Msol} from the pre-main sequence evolution tracks of \cite{Baraffe2015} at $\sim$\SI{e5}{\years} of evolution (their first recorded snapshot). We kept the overall disk structure constant for all setups with the values from \cref{tab:StarAndDiskParameters} and set the disk mass to $0.05\,M_*$.
We ran the simulations for all stellar masses once with dust diffusion and a diffusion parameter of $\delta=10^{-3}$ and once without dust diffusion. Gas evolution was turned off in these simulations, as we were only interested in the accuracy of our dust evolution model.
We present three of the simulations with diffusion in \cref{fig:StellarMasses}, where the first two panels in each row show snapshots of the simulations. The respective \dpy{} model is shown on top and the \tpop{} results below. We found that the overall size distribution evolution is well reproduced for most of the simulation domain and runtime. Our model is also able to capture the shape of the size distributions, as can be seen in the right panels. In the early stages of dust evolution (left panel of each row in \cref{fig:StellarMasses}), the dust size distribution is in coagulation-fragmentation equilibrium, which results in a typical power-law shape of the distribution with exponent $p=-3.5$. 
In the case of the \SI{0.1}{\Msol} star system, the dust is still in its initial growth phase at the time of the earlier snapshot. Therefore, the distribution has not yet fully reached the equilibrium state and is still more top-heavy---which is typical for the growth phase in which smaller particles are swept up by bigger particles.
In this phase, the distribution is not so well reproduced by a power law. Our three-parameter model nonetheless captures the steeper slope well around the size distribution peak, which contains most of the mass.

\begin{figure*}[t]
    \includegraphics[width=\textwidth]{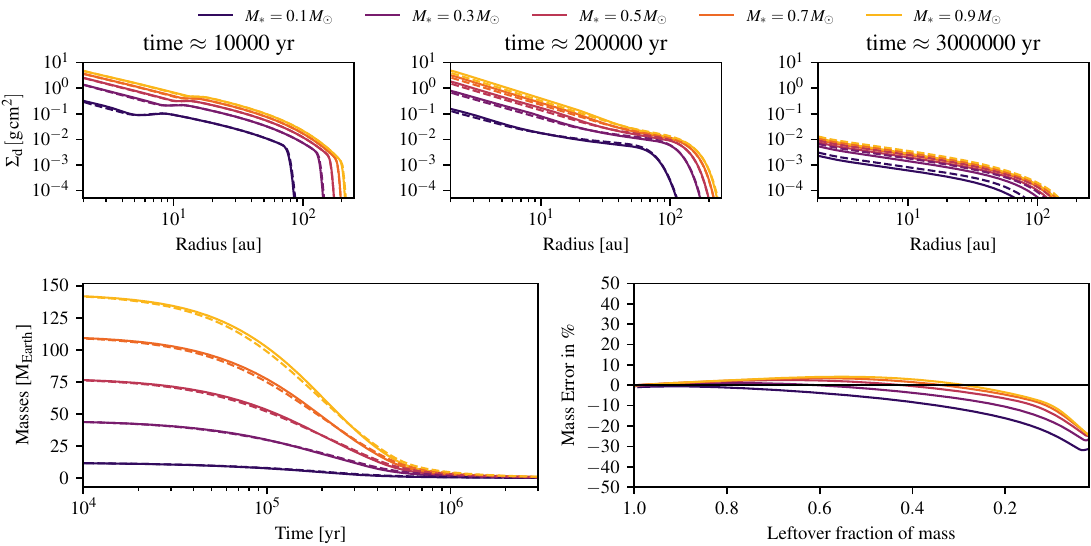}
    \caption[Timeseries comparison between models with different stellar masses.]{Comparison between \dpy{} and our model in setups with different stellar masses. \revI{Solid lines show the results of our \tpop{} simulations and dashed lines show the respective \dpy{} simulations.} The upper row shows a time series of the dust column density evolution in three snapshots. In the lower row, we show the mass evolution and the errors with respect to the full coagulation model \dpy{}.}
    \label{fig:StellarMasses_Diff}    
\end{figure*}

We present a more detailed look at the column density and mass evolution in \cref{fig:StellarMasses_Diff} (simulations without diffusion can be found in \cref{fig:StellarMasses_NoDiff}). 
The relative mass error with respect to the full coagulation model is always $\lesssim$\SI{20}{\percent} until $\sim$\SI{90}{\percent} of the mass has been accreted. The small errors at the beginning of the simulations are seen to add up in the late stages of disk evolution, where the absolute relative mass error in simulations is $\lesssim$\SI{35}{\percent} at the end of the simulation when more than $\SI{99}{\percent}$ of the mass has drifted out of the simulation domain after \SI{3e6}{\years}.

For stellar masses above \SI{0.3}{\Msol}, the errors are even smaller. Here, we determine a deviation of less than \SI{10}{\percent} until $\sim$\SI{90}{\percent} of the dust mass has drifted out of the simulation domain. Afterward, in the final stages of dust mass evolution, absolute errors increase again and are generally <\SI{30}{\percent}. Only for the system with the least massive star, we measured a maximum deviation of $\sim$\SI{35}{\percent} at the very end of the dust mass evolution.
One reason for this could be the smaller disk size, which means the dust growth front reaches the outer edge of the disk earlier. This area of strong radial dust-to-gas ratio gradients seems to be the origin of most of the deviations, as can be seen in the upper right panel of \cref{fig:StellarMasses_Diff}. 
Overall, the mass evolution in disks around stars of various masses is very well reproduced.

\revI{
\subsection{Varying fragmentation velocity}
To assess the accuracy of our model for different fragmentation velocities we chose a simple setup with a temperature-dependent fragmentation velocity. Following\footnote{See \url{https://stammler.github.io/dustpy/example_ice_lines.html} for the corresponding \dpy{} setup.}
\begin{equation}
    v_\mathrm{frag}=
    \begin{cases}
        \SI{100}{\centi \meter \per \second} &\text{ for }  T >\SI{150}{\kelvin} \text{ (ices evaporated)}\\
        \SI{1000}{\centi \meter \per \second} &\text{ for } \SI{150}{\kelvin}> T >\SI{80}{\kelvin} \text{ (H$_2$O ice)}\\
        \SI{700}{\centi \meter \per \second} &\text{ for } \SI{80}{\kelvin}> T >\SI{44}{\kelvin} \text{ (NH$_3$ ice)}\\   
        \SI{100}{\centi \meter \per \second} &\text{ for } \SI{44}{\kelvin}> T \text{ (CO$_2$ ice)}, \\ 
    \end{cases}
    \label{eq:vfrag_vary}
\end{equation}
we assumed particles covered in water ice to be the stickiest, followed by NH$_3$ ice and bare silicates or CO$_2$ ice. We used this highly simplified model only to illustrate that \tpop{} accurately reproduces the \dpy{} results for setups with different fragmentation velocities.
We display the resulting dust size distributions for two snapshots of the simulations in \cref{fig:icelines}. In the disk regions within $\sim$$\SI{2}{\AU}$, temperatures exceed \SI{150}{\kelvin} and the fragmentation velocity is accordingly low since all ices are assumed to be evaporated. Thus particles only grow up to millimeter sizes. In the regions beyond the water snow line, we find the two regions with larger fragmentation velocities; first the region with water ice, then, outside of $\sim$$\SI{8}{\AU}$ the region with ammonia ice which we assigned a slightly lower fragmentation velocity. The outer regions of the disk beyond the CO$_2$ snow line have again been given the lowest fragmentation velocity and have thus also lower maximum particle sizes.
As can be seen in \cref{fig:icelines}, the overall evolution of the disk and the local dust size distributions in the distinct regions of our \dpy{} and \tpop{} models show excellent agreements in dust column densities and maximum particle sizes.
\begin{figure*}
    \includegraphics[width=\textwidth]{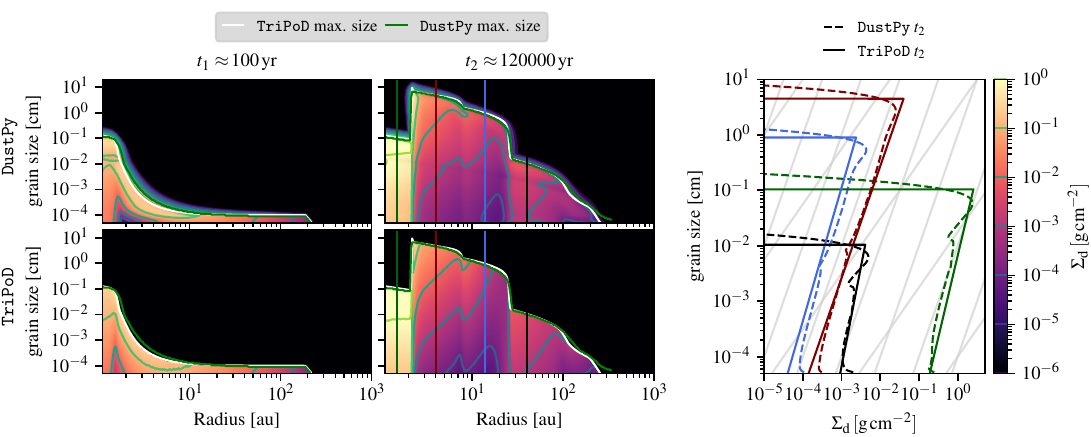}
    \caption{\revI{Comparison between a \dpy{} simulation and a \pluto{} simulation with our \tpop{} dust coagulation model of a protoplanetary disk with radially varying fragmentation velocity (see \cref{eq:vfrag_vary}). The left and center columns depict the global dust distribution at two different snapshots of the simulations, with the \dpy{} simulation in the top row and the \tpop{} simulation at the bottom. The right plot shows the local dust size distributions in the second snapshot at the locations of the vertical lines of the same color in the central panel. The gray lines in the right panel follow power laws with $q=-3.5$ and $q=-3.0$.}}
    \label{fig:icelines}
\end{figure*}
}

\subsection{Planetary gap}
\begin{figure*}[t]
    \centering
    \includegraphics[width=\textwidth]{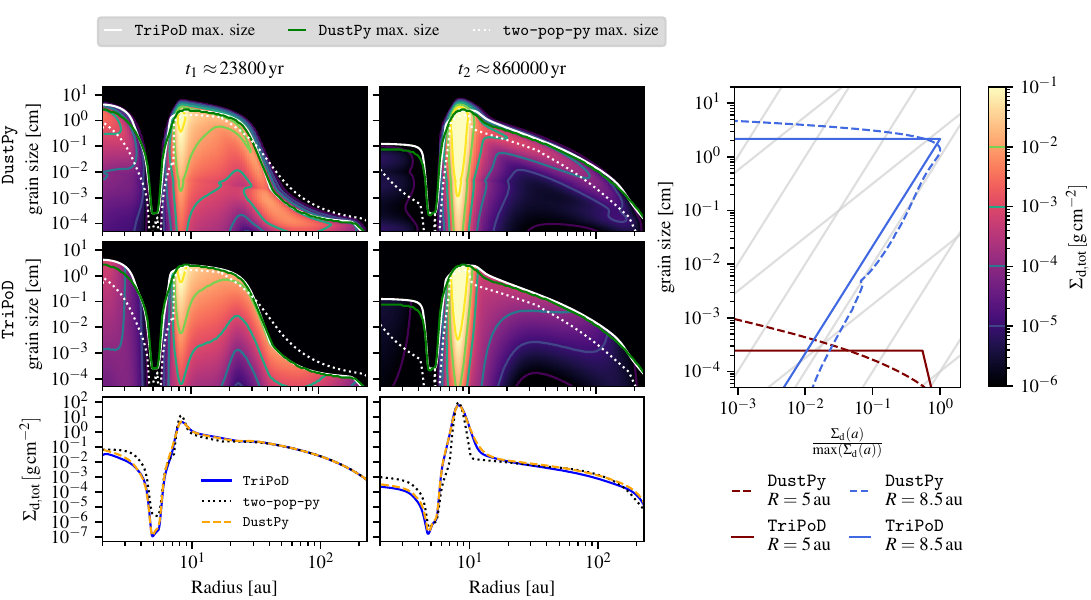}
    \caption[Detailed comparison between \dpy{} and \tpop{} in a simulation with a planetary gap.]{Comparison between a \dpy{} simulation and a \pluto{} simulation with our \tpop{} dust coagulation model in a protoplanetary disk with a pre-defined gap. We also show the particle sizes and dust column density profiles calculated with the old \tpoppy{} model as dotted lines. \revI{The plot on the right-hand side depicts the local dust size distributions at \SI{5}{\AU} (within the gap) and at \SI{8.5}{\AU} (at the pressure maximum), for two different snapshots. The gray lines in the right panel follow power laws with $q=-3.5$ and $q=-3.0$.}}
    \label{fig:Duffell1D}
\end{figure*}

\rev{As a next test case, we conducted a simulation with a planet-induced gap. For the gap in our one-dimensional \dpy{} and \tpop{} simulations we used the same gap profile as in \cref{sec:GapCalib}. Instead of letting the disk viscously evolve, we again turned off gas evolution for these simulations and imposed the gap profile immediately on the disk's column density structure.
We compare the resulting size distributions in \cref{fig:Duffell1D} with a similar \dpy{} model in the top row. As can be seen, dust particles collect in the pressure maximum outside of the planetary gap. The gap itself becomes depleted of dust, as the particles drift out and cannot be replenished due to the effect of the pressure bump. As a result, the maximum particle size in the full coagulation model is reduced within the gap.} As large grains are removed, the maximum particle size is also reduced in \tpop{}, and thus a reasonable size distribution is retained. In that way, a good fit with the particle size in the gap of the full coagulation model is achieved.
The size distribution in the full coagulation model is no longer top-heavy inside the gap but declines toward smaller sizes. 
The same is achieved in \tpop{} with our size reduction method, as seen in the right panels of \cref{fig:Duffell1D}. 
Particles collecting in the pressure bump outside of the gap are fragmentation-limited and can thus grow to larger sizes than in the outer regions, which are drift-limited. Most mass from the outer disk collects around the pressure maximum and does not reach the inner disk. 

The size distribution in the regions inside of the gap becomes drift-limited and dust densities decrease significantly due to the decreased inflow from the outer regions. 
Only diffusion of small particles through the gap maintains some particle flow in the inner disk.
For comparison, we also show the dust column densities and maximum particle sizes calculated with the old \tpoppy{} method in \cref{fig:Duffell1D}.
As can be seen, the new method agrees much better with \dpy{}. In the \tpoppy{} simulation, we find a much narrower peak in the dust densities around the pressure maximum due to the lack of a small particle population. Grain sizes are also underestimated in the gap, as \tpoppy{} sets the maximum particle size to the drift limit. This assumes an equilibrium between transport and coagulation which is not given in the gap, where the present particles are diffusing in from the outer edge and large particles are quickly removed by drift.

\subsection{Two-dimensional planet-disk simulation}
As a last test case, we ran a two-dimensional simulation of a protoplanetary disk with a Jupiter-mass planet. We set up a simulation in polar coordinates. The domain is spanning \SIrange{4}{34}{\AU} radially and full $2\pi$ azimuthally at a resolution of 1024 cells in the radial and azimuthal direction. The planet is represented by an additional gravitational potential following 
\begin{equation}
    \Phi_\mathrm{p}=-\frac{G M_\mathrm{p}}{\sqrt{d^2+r_\mathrm{sm}^2}}\, ,
\end{equation}
where $M_\mathrm{p}$ is the planets mass, $d$ is the distance to the planet, and $r_\mathrm{sm}=0.7\,H$ is the \revI{gravitational} smoothing length. The full gravitational potential in our simulation domain is then given as
\begin{equation}
    \Phi_\mathrm{tot} = \Phi_* + \Phi_\mathrm{p}\, ,
\end{equation}
where $\Phi_*$ is the gravitational potential of a Solar-mass star.
The disk was set up as a radial power law in column density and temperature with an isothermal equation of state. Details on the disk structure and simulation setup can be found in \cref{tab:2Dparams}.
We employed a viscosity $\nu=\alpha c_\mathrm{s}H$, with $\alpha=10^{-3}$. 
The hydrodynamic equations were solved with the HLL Riemann solver, using the third-order accurate Runge-Kutta scheme for the time integration and piece-wise-polynomial spatial reconstruction scheme to the fifth-order.

For the gas velocities, we used a zero-gradient boundary condition at the inner boundary, where we kept the gas density and the azimuthal velocity fixed to the initial values in the ghost cells. Similar boundary conditions were applied at the outer domain edge, with the difference that we allowed for outgoing velocities but not for inflow. The dust densities at the outer boundary were also kept at the initial value for 1000 orbital time scales after which we began to decrease them exponentially on a timescale of 1000 orbits in order to simulate a reduced dust inflow.

\begin{figure*}[t]
    \centering
    \includegraphics[width=\textwidth]{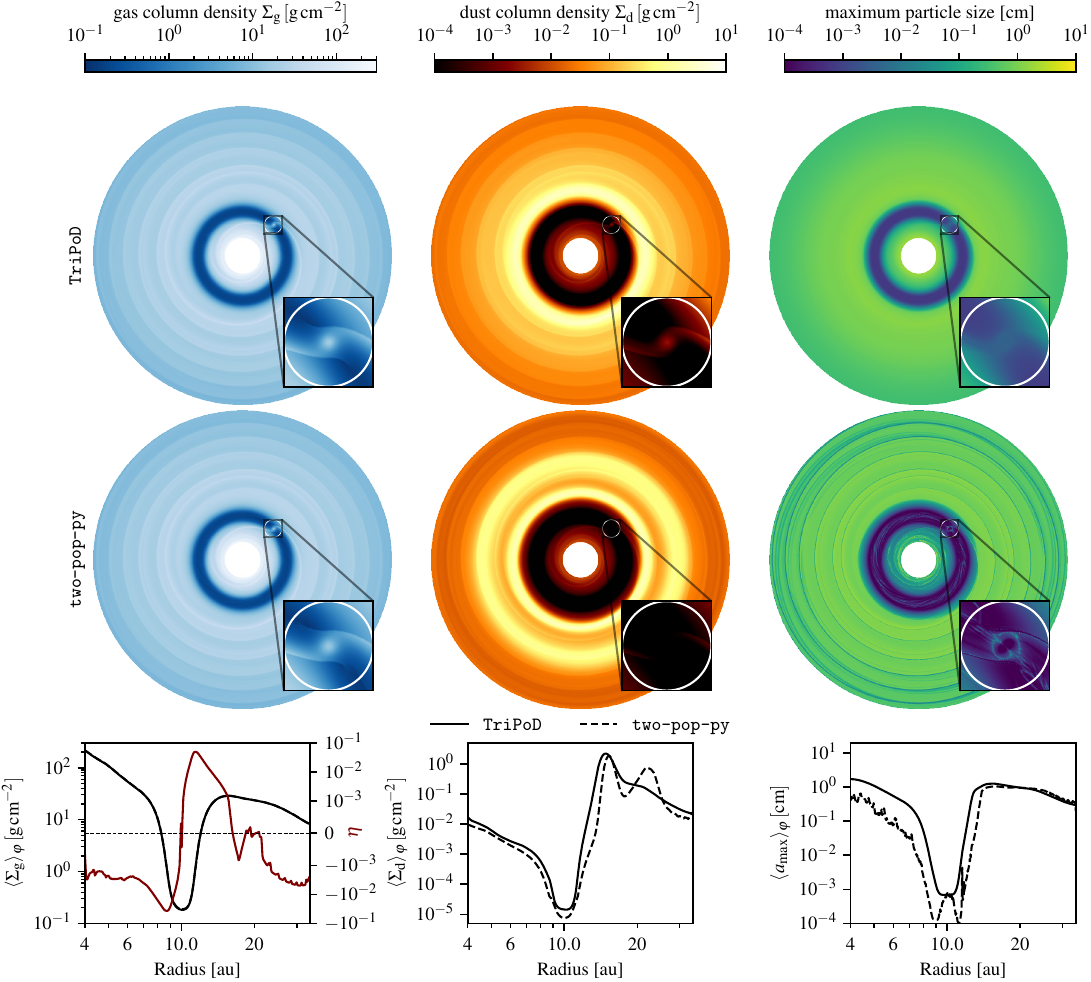}
    \caption[Two-dimensional simulations with a Jupiter-mass planet with \texttt{two-pop-py} and the new \tpop{} model.]{Comparison between two-dimensional \pluto{} simulations with the \tpop{} coagulation model and with the \tpoppy{} model. \revI{Both simulations have the exact same gas density structure and are only distinguished by the applied dust coagulation model.} The lower panels show the azimuthally averaged quantities. For the azimuthal averages, we have masked the region around the planet that has been marked with the white circle. \revI{The red line in the left panel shows the radial pressure gradient parameter $\eta=\frac{1}{2}\left(\frac{H}{R}\right)^2\frac{\partial \log{P}}{\partial \log{R}}$. It can be seen that in our simulation a weak second pressure maximum appears at $\sim$$\SI{20}{\AU}$, which leads to strong trapping in the \tpoppy{} model \citep{Birnstiel2012}. This effect is much less pronounced if a polydisperse size distribution is considered as done in our \tpop{} simulation, where the small dust population is less efficiently trapped.}}
    \label{fig:2DSim}
\end{figure*}

We compare the results to a simulation with the \tpoppy{} model, which only has one dust fluid.
The comparison is presented in \cref{fig:2DSim}, where the top row shows the \tpop{} simulation, and the middle row shows \tpoppy{} after 2000 planetary orbits of evolution. 
In the lower panels, we present the respective quantities' azimuthal averages, for which we have masked out the region around the planet that has been marked with a white circle.

The different grain sizes and the redistribution of mass because of fragmentation and coagulation between the two populations lead to different dust density distributions in the simulation with \tpop{} compared to the simulation with \tpoppy{}. 
The grains in the \tpoppy{} simulation collect in two narrowly peaked overdensities at the pressure maximum outside of the planetary gap and a weak second pressure perturbation. The gap is more depleted than in the simulation with the \tpop{} model. Furthermore, dust densities in \tpoppy{} are strongly enhanced in the weak second pressure perturbation that has almost no visible effect in the simulation with \tpop{}. The reason for this is the absence of a smaller, separately advected grain population in \tpoppy{} for which the effect of trapping would be much weaker. 
This is accounted for in the \tpop{} simulation, where the smaller dust population broadens the dust peaks significantly due to the weaker trapping of small grains. This also has the effect that more dust can diffuse through the gap, which means that the densities inside and in the gap itself are \revII{slightly} higher in \tpop{} than in \tpoppy{}.

Inside the gap, \tpop{} limits the size distribution exponent to a minimum value of $q\approx 4$ due to the applied size reduction rates which means the distribution is no longer fragmentation-limited but dominated by transport effects. The particles remaining in the gap are accordingly small and the size distributions are no longer top-heavy, as seen in the work by \cite{Drazkowska2019}. 
The \tpoppy{} model instead assumes the drift limit in the gap and inside of it. \revII{The size distribution, assumed by \texttt{two-pop-py} is thus vastly different from the almost flat or declining distribution seen in \tpop{} and in full coagulation models. Similar to the results of \cite{Drazkowska2019}, we find that the \texttt{two-pop-py} method still approximates the dust column densities inside of the gap fairly well. The Dust grain sizes are, however, underestimated inside and around the gap and the density profile at the pressure trap is not well reproduced by \texttt{two-pop-py}.} 

Our results are thus qualitatively similar to the conclusions by \cite{Drazkowska2019}. Coagulation makes it possible for small grains to pass through the gap via diffusion and coagulate again inside of the planet's orbit. 
The dust accumulates in the pressure bump, but the peak in the dust density is broader than in a simulation without coagulation, due to the presence of small grains, which are constantly produced in our simulation as a result of fragmentation.

\begin{table}[t]
\caption{Initial conditions for our two-dimensional \pluto{} simulation with \tpop{}.}
    \centering
    \begin{tabular*}{\columnwidth}{@{\extracolsep{\fill}}lr}
    \toprule 
    Simulation Parameter & Value \\
    \midrule
    \midrule
        Gas viscosity parameter ($\alpha$) & $10^{-3}$ \\
        \midrule
        Gas column density at \SI{1}{\AU} & \SI{1700}{\gram \per \square \centi \meter} \\
        Gas column density exponent ($\beta_\Sigma$) & -1.5 \\
        \midrule
        Temperature at \SI{1}{\AU} & \SI{195}{\kelvin} $\Leftrightarrow \nicefrac{H}{R}\big|_{\SI{10}{\AU}}=0.05$ \\
        Temperature exponent ($\beta_T$) & -0.5 \\
        \midrule
        Planet mass & \SI{1}{M_\mathrm{Jupiter}} \\
        Planet-star distance (circular orbit) & \SI{10}{\AU} \\
        \midrule
        Dust-to-gas ratio ($\varepsilon_\mathrm{tot}$) & 0.01 \\
        Initial dust size ($a_\mathrm{ini}$) & \SI{e-4}{\centi \meter} \\
        Minimum dust size ($\amin$) & \SI{e-5}{\centi \meter} \\
        Dust material density ($\rho_\mathrm{m}$) & \SI{1.2}{\gram \per \cubic \centi \meter} \\
        Dust fragmentation velocity ($v_\mathrm{frag}$) & \SI{1000}{\centi \meter \per \second} \\
        Dust diffusion parameter ($\delta$)  & $10^{-3}$ \\
        \bottomrule
    \end{tabular*}
    \tablefoot{Identical to the initial conditions in \cite{Drazkowska2019}.}
    \label{tab:2Dparams}
\end{table}

\section{Discussion}
\label{sec:discussion}
In its current form in the \pluto{} code, our model operates in the terminal velocity approximation. The dust fluids are implemented as passive tracers with modified fluxes. Thus, feedback from the dust on the gas is not included in our two-dimensional test simulation. However, as shown by \cite{Drazkowska2019}, the dust feedback in their two-dimensional planet-disk simulation only had a minor effect on the simulation outcome.
Effects like the streaming instability \citep{Youdin2005}, cannot be simulated with our current version of \tpop{} due to this limitation of our very simple dust fluid implementation. Unfortunately, \pluto{} does currently not have a multi-fluid feature.  A future iteration of the model could be combined with a code that supports dust fluids that are treated more self-consistently, like \texttt{FARGO3D} \citep{Benitez-Llambay2016,Benitez-Llambay2019} or \texttt{Athena++} \citep{Stone2020,Huang2022}. 
The terminal velocity approach also means that effects occurring on timescales shorter than the friction timescale of the particles cannot be modeled in the current form of \tpop{}. Possible cases could be the effects of shock waves, or spiral density waves.

Another shortcoming arises from the way we calculate the dust fluxes. As the maximum particle size changes throughout a simulation domain, neighboring cells usually do not have the same particle sizes. Therefore, the bin interfaces separating $\Sigma_0$ and $\Sigma_1$ are also different between neighboring cells. Our advection scheme does not take this into account in its current form, where transport is only occurring between the same size bin and not across bins. 
Due to the smooth variation of $\amax$ throughout simulations of protoplanetary disks, neighboring cells will typically still have very similar maximum particle sizes. The effect of this inaccuracy in the advection scheme is therefore likely small. 
However, one could construct extreme cases in which the error would be large, for example, if two neighboring cells had vastly different maximum particle sizes. 
We will have to address this issue in a future version of the model.

Details of the grain size distribution that are modeled in full-fledged coagulation models like \dpy{} cannot be reproduced in our simplified prescription. To still achieve a good fit with such models, our model has several calibration factors that have been adjusted to reach a good fit with \dpy{}.
However, this treatment of dust coagulation cannot reproduce the finer details of the coagulation process.

\rev{Our model omits the effects of bouncing \citep[see][]{Dominik2024}. This effect can lead to a steep, top-heavy size distribution and smaller maximum particle sizes. Bouncing could be implemented in a future version of \tpop{}.}

\rev{\tpop{}, as well as \dpy{} are designed to simulate the dust evolution in vertically integrated models of protoplanetary disks, assuming settling-mixing equilibrium at all times. Effects like sedimentation-driven coagulation \citep{Zsom2011,Krijt2016} can thus not be modeled by \dpy{} or \tpop{} in its current form. This is however not a fundamental limitation of \tpop{}, which could be adapted to work in three-dimensional setups in a future version.}
% In its current vertically integrated form, . Furthermore, our vertically integrated prescription cannot take three-dimensional turbulence, or vertically varying diffusivities into account.
% We are planning to develop a fully-three-dimensional version of the model that can consider these effects. 

\section{Summary and conclusion}
\label{sec:summary}

We present \tpop{}, an accurate and computationally inexpensive sub-grid model for dust coagulation in vertically integrated hydrodynamic simulations of protoplanetary disks. 
\tpop{} only requires two dust fluids and a tracer fluid representing the maximum particle size to model the evolution of a polydisperse dust size distribution. This makes it possible to run inexpensive simulations of planet-disk systems, protoplanetary disks with vortices, etc., and deduce the particle properties and dust densities.

We have implemented our model in the \pluto{} code. The workflow of the model during one timestep of a hydrodynamic simulation is as follows:
\begin{enumerate}
    \item \pluto{} calculates the gas fluxes across the cell interfaces with one of the standard Riemann solvers and reconstruction methods that are provided with the code.
        
    \item The dust fluxes and the particle size flux are calculated (\cref{eq:dr_flux} to \cref{eq:diff_flux}). Flux velocities are calculated from interpolated interface values. The drift flux is then taken upstream depending on the limited dust drift velocity. For dust diffusion, we calculate the limited diffusion flux in a manner identical to the \dpy{} model. The diffusion flux is then added to the total dust flux.
    (modifications in the \pluto{} code made in file $\texttt{adv\_flux.c}$)

    \new{
    \item The relative particle velocity $\Delta v_\mathrm{max}$ is calculated for the particle sizes $\amax$ and $f_{\Delta v}\amax$. It is used to determine the particle growth rate (\cref{eq:growthrate}).
    }
    \item The source terms for the large and the small dust fluids, are calculated. These fragmentation and sweep-up rates (\cref{eq:rate01,eq:rate10}) are then added to the source terms. \newline
If the large dust population is depleted to less than $f_\mathrm{crit}\varepsilon_\mathrm{tot}$, we calculate the respective dust depletion time and add the size reduction terms to the source terms in a way that conserves the current power law (\cref{eq:Shrink}). Mass is shifted accordingly from the small to the large population (\cref{eq:ShrinkMass}).
    The source terms of the two dust populations are calculated according to (\cref{eq:redistr}) and added to the equations' right-hand side (modification in \pluto{} made in file $\texttt{rhs\_source.c}\, $). 

    \item The array of conserved quantities is advanced by one timestep by \pluto{} using a standard time integration scheme provided with the code.
\end{enumerate}

Although \tpop{} is a highly simplified model, it can predict the dust mass evolution accurately for millions of years of evolution and simulate the effects of disk sub-structures on the dust size distribution, as demonstrated in our tests and calibration runs. Applications of this first version of \tpop{} could include:
\begin{itemize}
    \item More accurate studies of chemical networks in protoplanetary disks, which are highly dependent on the available grain surface area.
    \item Better radiative transfer post-processing of simulations, given the knowledge of the grain size distribution.
    \item Parameter studies of planet-disk systems with dust coagulation, that were so far infeasible due to the high computational cost of full coagulation models. 
    \item Simulations with self-consistently calculated thermal relaxation times.
\end{itemize}
In the future, we plan to extend this first version of the model to three-dimensional simulations of protoplanetary disks.

\section*{Acknowledgments}
T.P., H.K., and T.B. acknowledge the support of the German Science Foundation (DFG) priority program SPP 1992 “Exploring the Diversity of Extrasolar Planets” under grant Nos.\ BI 1816/7-2 and KL 1469/16-1/2. 
T.B. acknowledges funding from the European Research Council (ERC) under the European Union’s Horizon 2020 research and innovation program under grant agreement No 714769 and funding by the Deutsche Forschungsgemeinschaft (DFG, German Research Foundation) under grants 361140270, 325594231, and Germany’s Excellence Strategy - EXC-2094 - 390783311.
Computations were conducted on the computational facilities of the University of Munich (LMU).
We thank the anonymous referee for their constructive comments, which helped us to improve the quality of this article.

\bibliographystyle{aa} % style aa.bst
\bibliography{Literature}

\begin{thebibliography}{83}
\expandafter\ifx\csname natexlab\endcsname\relax\def\natexlab#1{#1}\fi

\bibitem[{Andrews {et~al.}(2018)Andrews, Huang, P{\'{e}}rez, Isella, Dullemond,
  Kurtovic, Guzm{\'{a}}n, Carpenter, Wilner, Zhang, Zhu, Birnstiel, Bai,
  Benisty, Hughes, {\"{O}}berg, \& Ricci}]{Andrews2018}
Andrews, S.~M., Huang, J., P{\'{e}}rez, L.~M., {et~al.} 2018, \apj, 869, L41

\bibitem[{Antonellini {et~al.}(2023)Antonellini, Kamp, \&
  Waters}]{Antonellini2023}
Antonellini, S., Kamp, I., \& Waters, L.~B. 2023, \aap, 672, A92

\bibitem[{Balbus \& Hawley(1991)}]{Balbus1991}
Balbus, S.~A. \& Hawley, J.~F. 1991, \apj, 376, 214

\bibitem[{Baraffe {et~al.}(2015)Baraffe, Homeier, Allard, \&
  Chabrier}]{Baraffe2015}
Baraffe, I., Homeier, D., Allard, F., \& Chabrier, G. 2015, \aap, 577

\bibitem[{Baruteau {et~al.}(2019)Baruteau, Barraza, P{\'{e}}rez, Casassus,
  Dong, Lyra, Marino, Christiaens, Zhu, Carmona, Debras, \&
  Alarcon}]{Baruteau2019}
Baruteau, C., Barraza, M., P{\'{e}}rez, S., {et~al.} 2019, \mnras, 486, 304

\bibitem[{Ben{\'{i}}tez-Llambay {et~al.}(2019)Ben{\'{i}}tez-Llambay, Krapp, \&
  Pessah}]{Benitez-Llambay2019}
Ben{\'{i}}tez-Llambay, P., Krapp, L., \& Pessah, M.~E. 2019, Astrophys. J.
  Suppl. Ser., 241, 25

\bibitem[{Ben{\'{i}}tez-Llambay \& Masset(2016)}]{Benitez-Llambay2016}
Ben{\'{i}}tez-Llambay, P. \& Masset, F.~S. 2016, Astrophys. J. Suppl. Ser.,
  223, 11

\bibitem[{Bergez-Casalou {et~al.}(2022)Bergez-Casalou, Bitsch, Kurtovic, \&
  Pinilla}]{Bergez-Casalou2022}
Bergez-Casalou, C., Bitsch, B., Kurtovic, N.~T., \& Pinilla, P. 2022, \aap,
  659, A6

\bibitem[{Binkert(2023)}]{Binkert2023}
Binkert, F. 2023, \mnras, 000, 1

\bibitem[{Birnstiel(2023)}]{Birnstiel2023}
Birnstiel, T. 2023, Annu. Rev. Astron. Astrophys. [\eprint[arXiv]{2312.13287}]

\bibitem[{Birnstiel {et~al.}(2009)Birnstiel, Dullemond, \&
  Brauer}]{Birnstiel2009}
Birnstiel, T., Dullemond, C.~P., \& Brauer, F. 2009, \aap, 503, 5

\bibitem[{Birnstiel {et~al.}(2010)Birnstiel, Dullemond, \&
  Brauer}]{Birnstiel2010}
Birnstiel, T., Dullemond, C.~P., \& Brauer, F. 2010, \aap, 513, 79

\bibitem[{Birnstiel {et~al.}(2018)Birnstiel, Dullemond, Zhu, Andrews, Bai,
  Wilner, Carpenter, Huang, Isella, Benisty, P{\'{e}}rez, \&
  Zhang}]{Birnstiel2018}
Birnstiel, T., Dullemond, C.~P., Zhu, Z., {et~al.} 2018, \apj, 869, L45

\bibitem[{Birnstiel {et~al.}(2012)Birnstiel, Klahr, \&
  Ercolano}]{Birnstiel2012}
Birnstiel, T., Klahr, H., \& Ercolano, B. 2012, \aap, 539, 148

\bibitem[{Birnstiel {et~al.}(2011)Birnstiel, Ormel, \&
  Dullemond}]{Birnstiel2011}
Birnstiel, T., Ormel, C.~W., \& Dullemond, C.~P. 2011, \aap, 525, 11

\bibitem[{Blandford \& Payne(1982)}]{Blandford1982}
Blandford, R.~D. \& Payne, D.~G. 1982, \mnras, 199, 883

\bibitem[{Brauer {et~al.}(2008)Brauer, Dullemond, \& Henning}]{Brauer2008}
Brauer, F., Dullemond, P., \& Henning, T. 2008, \aap, 480, 859

\bibitem[{Burn {et~al.}(2022)Burn, Emsenhuber, Weder, V{\"{o}}lkel, Klahr,
  Birnstiel, Ercolano, \& Mordasini}]{Burn2022}
Burn, R., Emsenhuber, A., Weder, J., {et~al.} 2022, \aap, 666

\bibitem[{Charnoz \& Taillifet(2012)}]{Charnoz2012}
Charnoz, S. \& Taillifet, E. 2012, \apj, 753, 119

\bibitem[{Dominik \& Dullemond(2024)}]{Dominik2024}
Dominik, C. \& Dullemond, C.~P. 2024, \aap, 682, A144

\bibitem[{Dr{\k{a}}{\.z}kowska {et~al.}(2023)Dr{\k{a}}{\.z}kowska, Bitsch,
  Lambrechts, Mulders, Harsono, Vazan, Liu, Ormel, Kretke, Morbidelli,
  Dr{\k{a}}{\.z}kowska, Bitsch, Lambrechts, Mulders, Harsono, Vazan, Liu,
  Ormel, Kretke, Morbidelli, Chevance, Krumholz, McLeod, Ostriker, Rosolowsky,
  Sternberg, Dr{\c{a}}{\.{z}}kowska, Bitsch, Lambrechts, Mulders, Harsono,
  Vazan, Liu, Ormel, Kretke, \& Morbidelli}]{Drazkowska2022}
Dr{\k{a}}{\.z}kowska, J., Bitsch, B., Lambrechts, M., {et~al.} 2023, in
  Protostars Planets VII, ed. S.-i. Inutsuka, Y.~Aikawa, T.~Muto, K.~Tomida, \&
  M.~Tamura, Vol. 534 (San Francisco: ASPCS), 717--759

\bibitem[{Dr{\k{a}}{\.z}kowska {et~al.}(2019)Dr{\k{a}}{\.z}kowska, Li,
  Birnstiel, Stammler, \& Li}]{Drazkowska2019}
Dr{\k{a}}{\.z}kowska, J., Li, S., Birnstiel, T., Stammler, S.~M., \& Li, H.
  2019, \apj, 885, 91

\bibitem[{Dr{\k{a}}{\.z}kowska {et~al.}(2021)Dr{\k{a}}{\.z}kowska, Stammler, \&
  Birnstiel}]{Drazkowska2021}
Dr{\k{a}}{\.z}kowska, J., Stammler, S.~M., \& Birnstiel, T. 2021, \aap, 647, 15

\bibitem[{Dubrulle {et~al.}(1995)Dubrulle, Morfill, \& Sterzik}]{Dubrulle1995}
Dubrulle, B., Morfill, G., \& Sterzik, M. 1995, \icarus, 114, 237

\bibitem[{Duffell(2020)}]{Duffell2020}
Duffell, P.~C. 2020, \apj, 889, 16

\bibitem[{Dullemond {et~al.}(2018)Dullemond, Birnstiel, Huang, Kurtovic,
  Andrews, Guzm{\'{a}}n, P{\'{e}}rez, Isella, Zhu, Benisty, Wilner, Bai,
  Carpenter, Zhang, \& Ricci}]{Dullemond2018}
Dullemond, C.~P., Birnstiel, T., Huang, J., {et~al.} 2018, \apj, 869, L46

\bibitem[{Emsenhuber {et~al.}(2021)Emsenhuber, Mordasini, Burn, Alibert, Benz,
  \& Asphaug}]{Emsenhuber2021}
Emsenhuber, A., Mordasini, C., Burn, R., {et~al.} 2021, \aap, 656, 69

\bibitem[{Epstein(1924)}]{Epstein1924}
Epstein, P.~S. 1924, Phys. Rev., 23, 710

\bibitem[{Estrada \& Cuzzi(2008)}]{Estrada2008}
Estrada, P.~R. \& Cuzzi, J.~N. 2008, \apj, 682, 515

\bibitem[{Fromang \& Nelson(2009)}]{Fromang2009}
Fromang, S. \& Nelson, R.~P. 2009, \aap, 496, 597

\bibitem[{G{\'{a}}rate {et~al.}(2021)G{\'{a}}rate, Delage, Stadler, Pinilla,
  Birnstiel, Stammler, Picogna, Ercolano, Franz, \& Lenz}]{Garate2021}
G{\'{a}}rate, M., Delage, T.~N., Stadler, J., {et~al.} 2021, \aap, 655, 18

\bibitem[{Guillet {et~al.}(2020)Guillet, Hennebelle, {Pineau Des For{\^{e}}ts},
  Marcowith, Commer{\c{c}}on, \& Marchand}]{Guillet2020}
Guillet, V., Hennebelle, P., {Pineau Des For{\^{e}}ts}, G., {et~al.} 2020,
  \aap, 643, 17

\bibitem[{G{\"{u}}ttler {et~al.}(2010)G{\"{u}}ttler, Blum, Zsom, Ormel, \&
  Dullemond}]{Guttler2010}
G{\"{u}}ttler, C., Blum, J., Zsom, A., Ormel, C.~W., \& Dullemond, C.~P. 2010,
  \aap, 513, A56

\bibitem[{Huang \& Bai(2022)}]{Huang2022}
Huang, P. \& Bai, X.-N. 2022, Astrophys. J. Suppl. Ser., 262, 11

\bibitem[{Kobayashi {et~al.}(2016)Kobayashi, Tanaka, \&
  Okuzumi}]{Kobayashi2016}
Kobayashi, H., Tanaka, H., \& Okuzumi, S. 2016, \apj, 817, 105

\bibitem[{Kolmogorov(1941)}]{Kolmogorov1941}
Kolmogorov, A.~N. 1941, Dokl. AN SSSR, 30, 299

\bibitem[{Kornet {et~al.}(2001)Kornet, Stepinski, \&
  R{\'{o}}zyczka}]{Kornet2001}
Kornet, K., Stepinski, T.~F., \& R{\'{o}}zyczka, M. 2001, \aap, 378, 180

\bibitem[{Krijt \& Ciesla(2016)}]{Krijt2016}
Krijt, S. \& Ciesla, F.~J. 2016, \apj, 822, 111

\bibitem[{Lau {et~al.}(2022)Lau, Dr{\k{a}}{\.z}kowska, Stammler, Birnstiel, \&
  Dullemond}]{Lau2022}
Lau, T. C.~H., Dr{\k{a}}{\.z}kowska, J., Stammler, S.~M., Birnstiel, T., \&
  Dullemond, C.~P. 2022, \aap, 668, 170

\bibitem[{Leiendecker {et~al.}(2022)Leiendecker, Jang-Condell, Turner, \&
  Myers}]{Leiendecker2022}
Leiendecker, H., Jang-Condell, H., Turner, N.~J., \& Myers, A.~D. 2022, \apj,
  941, 172

\bibitem[{Lenz {et~al.}(2020)Lenz, Klahr, Birnstiel, Kretke, \&
  Stammler}]{Lenz2020}
Lenz, C.~T., Klahr, H., Birnstiel, T., Kretke, K., \& Stammler, S. 2020, \aap,
  640

\bibitem[{Lesur {et~al.}(2023)Lesur, Ercolano, Flock, Lin, Yang, Barranco,
  Benitez-Llambay, Goodman, Johansen, Klahr, Laibe, Lyra, Marcus, Nelson,
  Squire, Simon, Turner, Umurhan, Youdin, Ercolano, Lin, Yang, Barranco,
  Benitez-Llambay, Goodman, Johansen, Klahr, Laibe, Lyra, Marcus, Nelson,
  Squire, Simon, Turner, Umurhan, Youdin, Lesur, Flock, Ercolano, Lin, Yang,
  Barranco, Benitez-Llambay, Goodman, Johansen, Klahr, Laibe, Lyra, Marcus,
  Nelson, Squire, Simon, Turner, Umurhan, \& Youdin}]{Lesur2022}
Lesur, G., Ercolano, B., Flock, M., {et~al.} 2023, in Protostars Planets VII,
  ed. S.-i. Inutsuka, Y.~Aikawa, T.~Muto, K.~Tomida, \& M.~Tamura, Vol. 534
  (San Francisco: ASPSC), 465--500

\bibitem[{Levermore \& Pomraning(1981)}]{Levermore1981}
Levermore, G.~D. \& Pomraning, G.~C. 1981, \apj, 248, 321

\bibitem[{Lichtenberg {et~al.}(2021)Lichtenberg, Dr{\k{a}}{\.z}kowska,
  Sch{\"{o}}nb{\"{a}}chler, Golabek, \& Hands}]{Lichtenberg2021}
Lichtenberg, T., Dr{\k{a}}{\.z}kowska, J., Sch{\"{o}}nb{\"{a}}chler, M.,
  Golabek, G.~J., \& Hands, T.~O. 2021, Science (80-. )., 371, 365

\bibitem[{Lombart {et~al.}(2024)Lombart, Br{\'{e}}hier, Hutchison, \&
  Lee}]{Lombart2024}
Lombart, M., Br{\'{e}}hier, C.-E., Hutchison, M., \& Lee, Y.-N. 2024, \mnras
  [\eprint[arXiv]{2404.11851}]

\bibitem[{Lombart {et~al.}(2022)Lombart, Hutchison, \& Lee}]{Lombart2022}
Lombart, M., Hutchison, M., \& Lee, Y.~N. 2022, \mnras, 517, 2012

\bibitem[{Lombart \& Laibe(2021)}]{Lombart2021}
Lombart, M. \& Laibe, G. 2021, \mnras, 501, 4298

\bibitem[{Marchand {et~al.}(2021)Marchand, Guillet, Lebreuilly, \& {Mac
  Low}}]{Marchand2021}
Marchand, P., Guillet, V., Lebreuilly, U., \& {Mac Low}, M.~M. 2021, \aap, 649,
  50

\bibitem[{Marchand {et~al.}(2022)Marchand, Guillet, Lebreuilly, \& {Mac
  Low}}]{Marchand2022}
Marchand, P., Guillet, V., Lebreuilly, U., \& {Mac Low}, M.~M. 2022, \aap, 666,
  27

\bibitem[{Marchand {et~al.}(2023)Marchand, Lebreuilly, {Mac Low}, \&
  Guillet}]{Marchand2023}
Marchand, P., Lebreuilly, U., {Mac Low}, M.~M., \& Guillet, V. 2023, \aap, 670,
  61

\bibitem[{Mathis {et~al.}(1977)Mathis, Rumpl, \& Nordsieck}]{Mathis1977}
Mathis, J.~S., Rumpl, W., \& Nordsieck, K.~H. 1977, \apj, 217, 425

\bibitem[{Mignone {et~al.}(2007)Mignone, Bodo, Massaglia, Matsakos, Tesileanu,
  Zanni, \& Ferrari}]{Mignone2007}
Mignone, A., Bodo, G., Massaglia, S., {et~al.} 2007, Astrophys. J. Suppl. Ser.,
  170, 228

\bibitem[{Muley {et~al.}(2023)Muley, Fuksman, \& Klahr}]{Muley2023}
Muley, D., Fuksman, J. D.~M., \& Klahr, H. 2023, \aap, 678, A162

\bibitem[{Nakagawa {et~al.}(1981)Nakagawa, Nakazawa, \& Hayashi}]{Nakagawa1981}
Nakagawa, Y., Nakazawa, K., \& Hayashi, C. 1981, \icarus, 45, 517

\bibitem[{Nakagawa {et~al.}(1986)Nakagawa, Sekiya, \& Hayashi}]{Nakagawa1986}
Nakagawa, Y., Sekiya, M., \& Hayashi, C. 1986, \icarus, 67, 375

\bibitem[{{\"{O}}berg {et~al.}(2021){\"{O}}berg, Guzm{\'{a}}n, Walsh, Aikawa,
  Bergin, Law, Loomis, Alarc{\'{o}}n, Andrews, Bae, Bergner, Boehler, Booth,
  Bosman, Calahan, Cataldi, Cleeves, Czekala, Furuya, Huang, Ilee, Kurtovic,
  {Le Gal}, Liu, Long, M{\'{e}}nard, Nomura, P{\'{e}}rez, Qi, Schwarz, Sierra,
  Teague, Tsukagoshi, Yamato, {van 't Hoff}, Waggoner, Wilner, \&
  Zhang}]{Oberg2021}
{\"{O}}berg, K.~I., Guzm{\'{a}}n, V.~V., Walsh, C., {et~al.} 2021, Astrophys.
  J. Suppl. Ser., 257, 1

\bibitem[{Ormel \& Cuzzi(2007)}]{Ormel2007}
Ormel, C.~W. \& Cuzzi, J.~N. 2007, \aap, 466, 413

\bibitem[{Ormel {et~al.}(2008)Ormel, Cuzzi, \& Tielens}]{Ormel2008}
Ormel, C.~W., Cuzzi, J.~N., \& Tielens, A. G. G.~M. 2008, \apj, 679, 1588

\bibitem[{Pascucci {et~al.}(2023)Pascucci, Cabrit, Edwards, Gorti, Gressel, \&
  Suzuki}]{Pascucci2022}
Pascucci, I., Cabrit, S., Edwards, S., {et~al.} 2023, in Protostars Planets
  VII, ed. S.-i. Inutsuka, Y.~Aikawa, T.~Muto, K.~Tomida, \& M.~Tamura (San
  Francisco: ASPCS), 567--604

\bibitem[{P{\'{e}}rez {et~al.}(2018)P{\'{e}}rez, Benisty, Andrews, Isella,
  Dullemond, Huang, Kurtovic, Guzm{\'{a}}n, Zhu, Birnstiel, Zhang, Carpenter,
  Wilner, Ricci, Bai, Weaver, \& {\"{O}}berg}]{Perez2018}
P{\'{e}}rez, L.~M., Benisty, M., Andrews, S.~M., {et~al.} 2018, \apj, 869, L50

\bibitem[{Pfeil {et~al.}(2023)Pfeil, Birnstiel, \& Klahr}]{Pfeil2023}
Pfeil, T., Birnstiel, T., \& Klahr, H. 2023, \apj, 959, 121

\bibitem[{Pfeil {et~al.}(2022)Pfeil, Cranmer, Ho, Armitage, Birnstiel, \&
  Klahr}]{Pfeil2022}
Pfeil, T., Cranmer, M., Ho, S., {et~al.} 2022, in Mach. Learn. Phys. Sci. Work.
  NeurIPS

\bibitem[{Pinilla {et~al.}(2021)Pinilla, Lenz, \& Stammler}]{Pinilla2021}
Pinilla, P., Lenz, C.~T., \& Stammler, S.~M. 2021, \aap, 645, 70

\bibitem[{Powell {et~al.}(2019)Powell, Murray-Clay, P{\'{e}}rez, Schlichting,
  \& Rosenthal}]{Powell2019}
Powell, D., Murray-Clay, R., P{\'{e}}rez, L.~M., Schlichting, H.~E., \&
  Rosenthal, M. 2019, \apj, 878, 116

\bibitem[{Schlichting \& Sari(2011)}]{Schlichting2011}
Schlichting, H.~E. \& Sari, R. 2011, \apj, 728, 68

\bibitem[{Simon {et~al.}(2022)Simon, Blum, Birnstiel, \&
  Nesvorn{\'{y}}}]{Simon2022}
Simon, J.~B., Blum, J., Birnstiel, T., \& Nesvorn{\'{y}}, D. 2022, Accept.
  Publ. Comets III [\eprint[arXiv]{2212.04509}]

\bibitem[{Smoluchowski(1916)}]{Smoluchowski1916}
Smoluchowski, M. 1916, Phys. Z., 17, 557

\bibitem[{Stammler \& Birnstiel(2022)}]{Stammler2022}
Stammler, S.~M. \& Birnstiel, T. 2022, \apj, 935, 35

\bibitem[{Stone {et~al.}(2020)Stone, Tomida, White, \& Felker}]{Stone2020}
Stone, J.~M., Tomida, K., White, C.~J., \& Felker, K.~G. 2020, Astrophys. J.
  Suppl. Ser., 249, 4

\bibitem[{Tamfal {et~al.}(2018)Tamfal, Dr{\k{a}}{\.z}kowska, Mayer, \&
  Surville}]{Tamfal2018}
Tamfal, T., Dr{\k{a}}{\.z}kowska, J., Mayer, L., \& Surville, C. 2018, \apj,
  863, 97

\bibitem[{Tsukagoshi {et~al.}(2022)Tsukagoshi, Nomura, Muto, Kawabe, Kanagawa,
  Okuzumi, Ida, Walsh, Millar, Takahashi, Hashimoto, Uyama, \&
  Tamura}]{Tsukagoshi2022}
Tsukagoshi, T., Nomura, H., Muto, T., {et~al.} 2022, \apj, 928, 49

\bibitem[{Tsukamoto \& Okuzumi(2022)}]{Tsukamoto2022}
Tsukamoto, Y. \& Okuzumi, S. 2022, \apj, 934, 88

\bibitem[{Vorobyov {et~al.}(2018)Vorobyov, Akimkin, Stoyanovskaya,
  Pavlyuchenkov, \& Liu}]{Vorobyov2018}
Vorobyov, E.~I., Akimkin, V., Stoyanovskaya, O., Pavlyuchenkov, Y., \& Liu,
  H.~B. 2018, \aap, 614, 98

\bibitem[{Vorobyov \& Elbakyan(2019)}]{Vorobyov2019a}
Vorobyov, E.~I. \& Elbakyan, V.~G. 2019, \aap, 631, A1

\bibitem[{Vorobyov {et~al.}(2020)Vorobyov, Matsukoba, Omukai, \&
  Guedel}]{Vorobyov2020}
Vorobyov, E.~I., Matsukoba, R., Omukai, K., \& Guedel, M. 2020, \aap, 638, A102

\bibitem[{Vorobyov {et~al.}(2019)Vorobyov, Skliarevskii, Elbakyan,
  Pavlyuchenkov, Akimkin, \& Guedel}]{Vorobyov2019}
Vorobyov, E.~I., Skliarevskii, A.~M., Elbakyan, V.~G., {et~al.} 2019, \aap,
  627, A154

\bibitem[{Weidenschilling(1977)}]{Weidenschilling1977}
Weidenschilling, S.~J. 1977, \mnras, 180, 57

\bibitem[{Weidenschilling(1980)}]{Weidenschilling1980}
Weidenschilling, S.~J. 1980, \icarus, 44, 172

\bibitem[{Wetherill \& Stewart(1989)}]{Wetherill1989}
Wetherill, G.~W. \& Stewart, G.~R. 1989, \icarus, 77, 330

\bibitem[{Whipple(1972)}]{Whipple1972}
Whipple, F. 1972, in From Plasma to Planet, ed. A.~Evlius (New York: Wiley
  Interscience Division), 211

\bibitem[{Youdin \& Goodman(2005)}]{Youdin2005}
Youdin, A.~N. \& Goodman, J. 2005, \apj, 620, 459

\bibitem[{Zsom {et~al.}(2011)Zsom, Ormel, Dullemond, \& Henning}]{Zsom2011}
Zsom, A., Ormel, C.~W., Dullemond, C.~P., \& Henning, T. 2011, \aap, 534, A73

\bibitem[{Zsom {et~al.}(2010)Zsom, Ormel, G{\"{u}}ttler, Blum, \&
  Dullemond}]{Zsom2010}
Zsom, A., Ormel, C.~W., G{\"{u}}ttler, C., Blum, J., \& Dullemond, C.~P. 2010,
  \aap, 513, A57

\end{thebibliography}

\begin{appendix}
\section{Additional tables}
\label{tab:Sym}
\begin{table}[!ht]
\small
\centering
\caption{Lists of symbols}
\resizebox{\textwidth}{!}{
    \begin{tabularx}{\columnwidth}[t]{>{$}l<{$}X}
\multicolumn{2}{>{\hsize=\dimexpr2\hsize+2\tabcolsep+\arrayrulewidth\relax}c}{\normalsize General symbols}\\
    \toprule
    \textup{Symbol} & Description \\
    \midrule
    \midrule 
       a   & particle size \\
       \dot{a} & dust growth rate \\
       a_\mathrm{drift} & drift-limited particle size \\
       a_\mathrm{drift\text{-}frag} & drift-fragmentation-limited particle size \\
       a_\mathrm{frag} & fragmentation-limited particle size \\
       a_\mathrm{mon}, a_\mathrm{gr} & monomer and grown dust size (\texttt{two-pop-py}) \\
       a_\mathrm{turb\text{-}frag} & turbulent-fragmentation-limited particle size \\
       c_\mathrm{s} & soundspeed \\
       D & dust diffusivity \\
       f_\mathrm{fric} & aerodynamic friction force \\
       f_\mathrm{m} & fudge factor (\texttt{two-pop-py}) \\
       G & gravitational constant \\
       H & gas scale height \\
       H_\mathrm{d} & dust scale height \\ 
       k_\mathrm{B} & Boltzmann constant \\
       L_* & stellar luminosity \\
       m & particle mass\\
       m_\mathrm{p} & proton mass \\
       M_* & stellar mass \\
       M_\mathrm{disk} & disk mass \\
       M_\odot & solar mass \\
       n(a), n(m) & number density size/mass distribution \\
       \mathcal{N} & size ratio for drift-induced collisions\\
       R & stellocentric cylindrical radius \\
       R_\mathrm{c} & characteristic disk radius \\
       \mathrm{Re} & Reynolds number \\
       P & gas pressure \\
       q & dust size distribution power-law exponent \\
       q_\mathrm{m} & dust mass distribution power-law exponent \\
       \St  & Stokes number \\
       t_\mathrm{fric} & stopping time \\
       t_\mathrm{grow} & dust growth time scale \\
       T & gas temperature \\
       \mathbb{T} & viscous stress tensor \\
       \overline{v} & average dust velocity (\texttt{two-pop-py}) \\
       v_{\text{d-g},R} & radial relative velocity of dust and gas \\
       v_\mathrm{frag} & dust fragmentation velocity \\
       v_\mathrm{g}, v_\mathrm{d} & gas and dust velocity \\
       v_\mathrm{K} &  Keplerian velocity \\
       v_\mathrm{mon}, v_\mathrm{gr} & monomer and grown dust vel. (\texttt{two-pop-py}) \\
       \Delta v & relative particle velocity (source in subscript) \\
       z & cylindrical distance from the disk midplane \\
       \midrule
       \alpha  & turbulence parameter \\
       \beta_\Sigma & double-log. column density gradient\\
       \beta_T & double-log. temperature gradient \\
       \gamma & abs.\ value of the double-log.\ pressure gradient \\
       \delta & dust diffusion parameter (here $\delta$=$\alpha$) \\
       \varepsilon & dust-to-gas ratio \\
       \lambda_\mathrm{mfp} & gas molecule mean free path \\
       \mu & mean molecular weight \\
       \nu & coagulation kernel index \\
       \nu_\mathrm{turb}, \nu_\mathrm{mol} & turbulent and molecular viscosity \\
       \xi & fragment size distribition \\
       \rhog, \rhod & gas and dust volume density \\
       \rho_\mathrm{g,mid}, \rho_\mathrm{d,mid} & gas and dust midplane volume density \\
       \rho_\mathrm{m} & material density of dust \\
       \sigma_\mathrm{H_2} & hydrogen molecule collision cross section \\
        \sigma_\mathrm{SB} & Stefan-Boltzmann constant \\
       \Omega, \OmK & angular frequency, Keplerian frequency \\
        \bottomrule
    \end{tabularx}
    \quad
    \begin{tabularx}{\columnwidth}[t]{>{$}l<{$}X}
    \multicolumn{2}{>{\hsize=\dimexpr2\hsize+2\tabcolsep+\arrayrulewidth\relax}c}{\normalsize Symbols specific to \tpop{}}\\
    \toprule
    \textup{Symbol} & Description \\
    \midrule
    \midrule
        \dot{a}_\mathrm{max} & growth rate of max.\ size \\
        a_0, a_1 & mass-averaged particle sizes of populations \\
        \amin, \amax & min.\ and max.\ size of distribution \\
        a_\mathrm{int} & $\sqrt{\amin \amax}$ \\
        a_\mathrm{lim} & lower limit for size reduction in gaps \\
        f_\mathrm{crit} & crit.\ mass depletion coefficient for shrinking\\
        f_\mathrm{drift} & drift velocity calibration factor \\
        f_{\Delta v} & collision speed parameter \\
\vspace{0.7mm} 
        {F}^\mathrm{drift}_{\Sigma_0, \, i+\nicefrac{1}{2}} & dust flux due to radial drift (small pop.) \\ \vspace{0.7mm}
        {F}^\mathrm{drift}_{\Sigma_1, \, i+\nicefrac{1}{2}} & dust flux due to radial drift (large pop.)\\ \vspace{0.7mm}
        {F}^\mathrm{drift}_{\amax\Sigma_1, \, i+\nicefrac{1}{2}} & flux of the density-weighted maximum particle size due to radial 
        drift\\ \vspace{0.7mm}
        {F}^\mathrm{diff}_{\Sigma_0, \, i+\nicefrac{1}{2}} & dust flux due to radial diffusion (small pop.) \\ \vspace{0.7mm}
        {F}^\mathrm{diff}_{\Sigma_1, \, i+\nicefrac{1}{2}} & dust flux due to radial diffusion (large pop.)\\ \vspace{0.7mm}
        {F}^\mathrm{diff}_{\amax\Sigma_1, \, i+\nicefrac{1}{2}} & flux of the density-weighted maximum particle size due to radial diffusion \\ \vspace{0.7mm}
        {F}_{\Sigma_{0/1},\, i\pm\nicefrac{1}{2}} & total flux of the dust fluids \\ \vspace{0.7mm}
        {F}_{\amax\Sigma_1, i\pm\nicefrac{1}{2}} & total flux of the density-weighted maximum particle size \\
        \mathcal{F} & size distribution calibration function \\
        \tilde{\mathcal{F}} & vert.\ int.\ size distribution calibration function \\
        H_0, H_1 & dust scale height of populations \\
        m_0, m_1 & particle masses corresponding to $a_0$ and $a_1$ \\
        N & number of bins of the reconstructed size distr. \\
        p_\mathrm{frag} & fragmentation/growth transition function \\
        p_\mathrm{sweep} &  $1-p_\mathrm{frag}$ \\
        p_\mathrm{turb} & turb.-dom./drift-dom. transition function \\
        p_\mathrm{drift} &  $1-p_\mathrm{turb}$ \\
        p_\mathrm{turb.1} & turb.1/turb.2 transition function\ \\
        p_\mathrm{turb.2} &  $1-p_\mathrm{turb.1}$ \\
        q_\mathrm{frag} & general fragmentation power-law exp.\ \\
        q_\mathrm{turb.1} & fragmentation power-law exp.\ in turb.\ 1 regime\\
        q_\mathrm{turb.2} & fragmentation power-law exp.\ in turb.\ 2 regime\\
        q_\text{drift-frag} & drift-fragmentation power-law exp.\ \\
        q_\mathrm{sweep} & non-equilibrium power-law exp.\ \\
        q_\text{turb-frag} & turbulent-fragmentation power-law exp.\ \\
        s & steepness parameter of the transition from growth to fragmentation \ \\
        {v}_{\Sigma_{0/1}} & drift flux velocity for populations \\
        \Delta v_{01} & rel.\ vel.\ between grains of sizes $a_0$ and $a_1$ \\
        \Delta v_{11} & rel.\ vel.\ between grains of sizes $a_1$ and $f_{\Delta v}a_1$ \\
        \Delta v_\mathrm{max} & rel.\ vel.\ between grains of sizes $\amax$ and $f_{\Delta v}\amax$ \\
        \midrule
        \varepsilon_0, \varepsilon_1 & vert.\ integrated dust-to-gas ratios of population \\
        \varepsilon_\mathrm{tot} & total dust-to-gas ratio \\
        \lambda_\mathrm{lim} & flux limiter \\
        \dot{\rho}_{0\rightarrow 1} & ``sweep-up'' rate \\
        \dot{\rho}_{1\rightarrow 0} & ``fragmentation'' rate \\
        \sigma_{01} & coll.\ cross sec.\ of grains of size $a_0$ and $a_1$  \\
        \sigma_{11} & coll.\ cross sec.\ of grains of size $a_1$ and $f_{\Delta v}a_1$  \\
        \Sigma_0, \Sigma_1 & dust column densities of populations \\
        \dot{\Sigma}_{0\rightarrow 1} & vertically integrated ``sweep-up'' rate \\
        \dot{\Sigma}_{1\rightarrow 0} & vertically integrated ``fragmentation'' rate \\
        \tau_\mathrm{depletion} & dust depletion timescale \\
        \bottomrule
    \end{tabularx}}
\end{table}
\clearpage

\rev{
\section{Column density formulation}
\label{app:VertInt}

Assuming $\sigma_{01,11}$, $\Delta v_{01,11}$, $m_{0,1}$, and $\mathcal{F}$ to be vertically constant we deduce the vertically integrated mass exchange rates (\cref{eq:rate01} and \cref{eq:rate10})
\begin{align}
    \dot{\Sigma}_{\text{d}, 0\rightarrow 1} &= \frac{\Sigma_0\Sigma_1 \sigma_{01}\Delta v_{01}}{m_1 2\pi H_0 H_1 } \int_{-\infty}^{\infty}\exp\left[-\frac{z^2}{2}\left(\frac{H_0^2+H_1^2}{H_0^2H_1^2}\right)\right] \, \mathrm{d}z \\
    &= \frac{\Sigma_0\Sigma_1 \sigma_{01}\Delta v_{01}}{ m_1\sqrt{2\pi (H_0^2+ H_1^2)}} \\
    \dot{\Sigma}_{\text{d}, 1\rightarrow 0} &= \frac{\Sigma_1^2 \sigma_{11}\Delta v_{11}}{m_1 2\pi H_1^2}\tilde{\mathcal{F}}  \int_{-\infty}^{\infty}\exp\left[-\frac{z^2}{H_1^2}\right] \, \mathrm{d}z \\
    &= \frac{\Sigma_1^2 \sigma_{11}\Delta v_{11}}{ m_1\sqrt{4\pi H_1^2}}\tilde{\mathcal{F}}
\end{align}
Thus, also $\mathcal{F}$ has a modified form in the vertically integrated setup
\begin{equation}
       \tilde{\mathcal{F}} =  \sqrt{\frac{2H_1^2}{H_0^2+H_1^2}}\frac{\sigma_{01}}{\sigma_{11}}\frac{\Delta v_{01}}{\Delta v_{11}}\left(\frac{\amax}{\aint}\right)^{- (q+4)}\,.
\end{equation}
}

\rev{
\section{Transition functions}
\label{sec:transfunc}
In order to determine the transition functions, we have taken into account the results of local dust coagulation simulations and theoretical models. The transitions should not be too rapid to not cause issues during the numerical integration. 

For the transition between the small regime of turbulence and the fully intermediate regime of turbulence we take the approximate  transition criterion from \cite{Ormel2008}
\begin{equation}
    f_\mathrm{turb.1}^\mathrm{turb.2} = \frac{5\, t_\mathrm{s}}{\tau_\mathrm{max}} 
    \begin{cases}
        >1 \quad \text{small particle regime (turb.1)} \\
        <1 \quad \text{fully inter. regime (turb.2), }
    \end{cases}
\end{equation}
where $\tau_\mathrm{max}$ is the friction time of the largest particles, and $t_\mathrm{s}= \Rey^{-\nicefrac{1}{2}} \OmK^{-1}$ is the small eddy turnover time.
Using the same functional form as for the growth rate \cref{eq:growthrate}, but now from 0 to 1 instead of -1 to 1, we define
\begin{equation}
    p_\mathrm{turb.1} \coloneqq \frac{1}{2}\left(1-\frac{\left(f_\mathrm{turb.1}^\mathrm{turb.2}\right)^4 -1}{\left(f_\mathrm{turb.1}^\mathrm{turb.2}\right)^4 + 1}\right).
\end{equation}
This expression is approaching 0 if $\tau_\mathrm{max}<5t_\mathrm{s}$ and 1 if $\tau_\mathrm{max}>5t_\mathrm{s}$.

We apply a very similar function to the transition from the drift-dominated regime to the turbulence-dominated regime.
We define 
\begin{equation}
    f_\mathrm{drift}^\mathrm{turb} = \frac{\Delta v_\mathrm{turb}}{\Delta v_\mathrm{drift}} 
    \begin{cases}
        >1 \quad \text{turbulence dominated} \\
        <1 \quad \text{drift dominated}
    \end{cases}\, ,
\end{equation}
where $\Delta v_\mathrm{turb}$ is the turbulent collision velocity. We define
\begin{equation}
    p_\mathrm{drift} \coloneqq \frac{1}{2}\left(1-\frac{\left(f_\mathrm{drift}^\mathrm{turb}\right)^6 -1}{\left(f_\mathrm{drift}^\mathrm{turb}\right)^6 + 1}\right)\, .
\end{equation}
where we found that a slightly steeper transition fits better with the results of local coagulation simulations.

Finally, we have to introduce the transition from growth to fragmentation. For this, we chose a steeper transition, following
\begin{equation}
    p_\mathrm{frag} = \exp\left[-\left(5\left(\min\left(\frac{\Delta v_\mathrm{tot}}{v_\mathrm{frag}},1.0\right)-1.0\right)\right)^2\right],
\end{equation}
which ensures that the equilibrium size distribution is reached on a fast enough timescale.
}

\rev{
\section{Flux-limited dust diffusion}
\label{sec:fluxlimiter}
In order to avoid unrealistically large diffusion fluxes in the presence of strong gradients in the dust-to-gas ratio, we introduce a flux limiter that is conceptually identical to the one by \cite{Levermore1981}, which was used by \cite{Stammler2022}.
Without the flux limiter, the diffusion flux is given by 
\begin{equation}
    \Vec{F}_\mathrm{diff} = -D \Sigmag \Vec{\nabla}\varepsilon \, .
\end{equation}
The transport velocity shall however not be larger than the turbulent velocities that drive the diffusion, meaning the maximum allowed flux is
\begin{equation}
    \Vec{F}_\mathrm{diff,max} \coloneqq v_\mathrm{turb}\varepsilon \Sigmag = \frac{\sqrt{\delta}c_\mathrm{s}}{1+\St^2}  \varepsilon \Sigmag \, .
\end{equation}
The flux limiter is now defined via
\begin{equation}
    \lambda_\mathrm{lim} = \frac{1+\chi}{1+\chi +\chi^2}\, ,
\end{equation}
where 
\begin{equation}
    \chi = \frac{\Vec{F}_\mathrm{diff}}{\Vec{F}_\mathrm{diff,max}}\, ,
\end{equation}
is the ratio of the diffusion flux and the maximum flux. The limited flux is then given by 
\begin{equation}
    \Vec{F}_\mathrm{diff, lim} =  \lambda_\mathrm{lim} \Vec{F}_\mathrm{diff} \, .
\end{equation}
Note that although the flux is a vector (with components for each direction), all operations here are component-wise. In \pluto{}, we define all quantities in the above equations at the cell interfaces.
}

\section{Calibrations and test simulations without diffusion}
Here, we present the same calibration and test simulations shown in the main part of this article, but now without dust diffusion ($\delta=0$).
\begin{figure*}[!ht]
    \includegraphics[width=\textwidth]{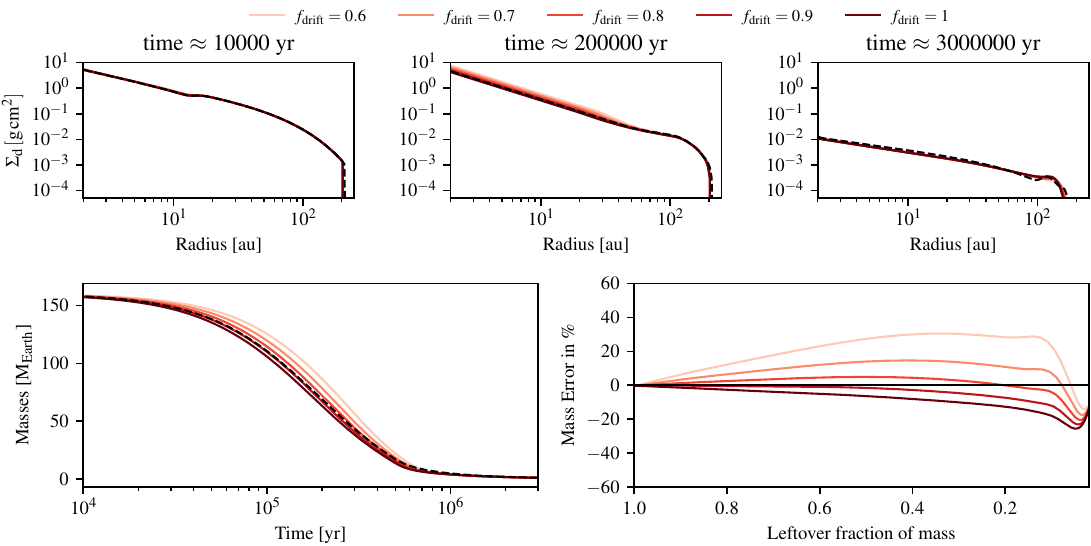}
    \caption[Timeseries comparison between models with different drift velocity factors without diffusion.]{Comparison between \dpy{} and our model in a setup without dust diffusion and with different drift calibration factors $f_\mathrm{drift}$. \revI{Solid lines show the results of our \tpop{} calibration runs and dashed lines show the respective \dpy{} simulation, to which we calibrate our model.} The upper row shows a timeseries of the dust column density evolution in three snapshots. In the lower row, we show the mass evolution and the errors with respect to the full coagulation model \dpy{}. For a factor of \new{$f_\mathrm{drift}=0.8$}, the mass evolution of the full coagulation model is well reproduced by our three-parameter model.}
    \label{fig:FudgeNoDiff}
\end{figure*}

\begin{figure*}[!ht]
    \includegraphics[width=\textwidth]{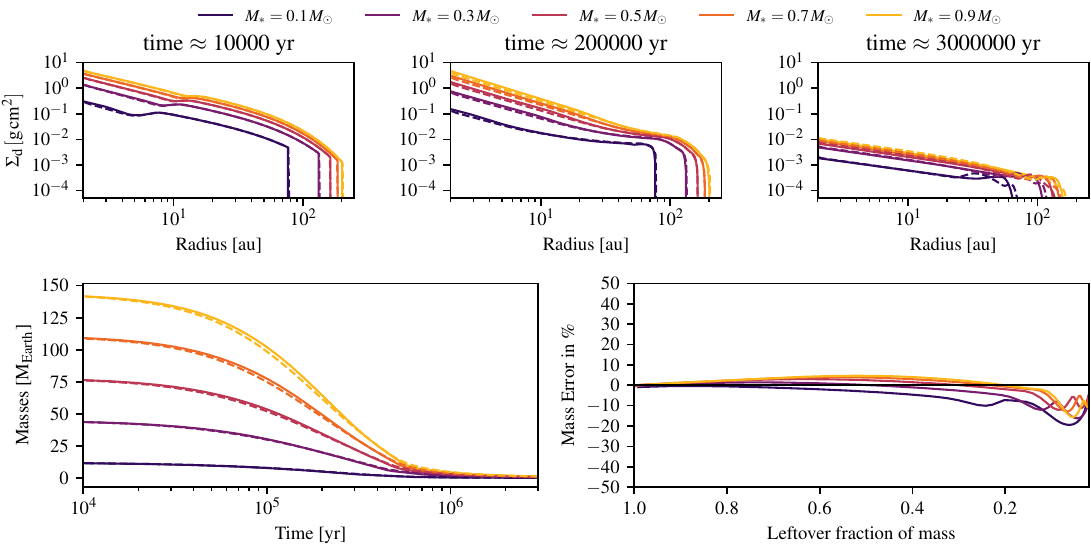}
\caption[Timeseries comparison between models with different stellar masses.]{Comparison between \dpy{} and our model in setups with different stellar masses without dust diffusion. \revI{Solid lines show the results of our \tpop{} simulations and dashed lines show the respective \dpy{} simulations.} The upper row shows a time series of the dust column density evolution in three snapshots. In the lower row, we show the mass evolution and the errors with respect to the full coagulation model \dpy{}.}
    \label{fig:StellarMasses_NoDiff}    
\end{figure*}
\end{appendix}

\end{document}